\documentclass[twocolumn,showpacs,preprintnumbers]{revtex4-2}
\usepackage{graphicx}
\usepackage{amsmath}
\newcommand{\be}{\begin{equation}}
\newcommand{\ee}{\end{equation}}
\newcommand{\bdm}{\begin{eqnarray}}
\newcommand{\edm}{\end{eqnarray}}
\newcommand{\kappafl}{\kappa_{\scriptscriptstyle F\!L}}
\begin{document}
\setlength{\abovedisplayskip}{6pt}
\setlength{\belowdisplayskip}{6pt}
\preprint{Accepted by The Astrophysical Journal}
\title{Perpendicular Diffusion of Energetic Particles: A Complete Analytical Theory}
\author{A. Shalchi}
\email{andreasm4@yahoo.com}
\affiliation{Department of Physics and Astronomy, University of Manitoba, Winnipeg, Manitoba R3T 2N2, Canada}
\date{\today}
\begin{abstract}
Over the past two decades scientists have achieved a significant improvement of our understanding of the transport of energetic
particles across a mean magnetic field. Due to test-particle simulations as well as powerful non-linear analytical tools our
understanding of this type of transport is almost complete. However, previously developed non-linear analytical theories do not
always agree perfectly with simulations. Therefore, a correction factor $a^2$ was incorporated into such theories with
the aim to balance out inaccuracies. In this paper a new analytical theory for perpendicular transport is presented. This theory
contains the previously developed unified non-linear transport theory, the most advanced theory to date, in the limit of small
Kubo number turbulence. For two-dimensional turbulence new results are obtained. In this case the new theory describes perpendicular
diffusion as a process which is sub-diffusive while particles follow magnetic field lines. Diffusion is restored as soon as
the turbulence transverse complexity becomes important. For long parallel mean free paths one finds that the perpendicular
diffusion coefficient is a reduced field line random walk limit. For short parallel mean free paths, on the other hand, one gets a
hybrid diffusion coefficient which is a mixture of collisionless Rechester \& Rosenbluth and fluid limits. Overall the new analytical
theory developed in the current paper is in agreement with heuristic arguments. Furthermore, the new theory agrees almost perfectly
with previously performed test-particle simulations without the need of the aforementioned correction factor $a^2$ or any other free
parameter.
\end{abstract}
\pacs{47.27.tb, 96.50.Ci, 96.50.Bh}
\maketitle
\section{Introduction}

A fundamental problem of modern space and astrophysics is the interaction between electrically charged energetic particles and
magnetic turbulence. These interactions are crucial in the theory of cosmic ray propagation as well as diffusive shock acceleration.
This concerns those phenomena in the solar system but also in other scenarios such as the interstellar media of our own and external
galaxies.

To understand and describe the interaction between turbulence and particles is difficult because of several reasons. First of all turbulence
itself is a very complicated topic due to the fact that it is a complex non-linear dynamical phenomenon (see Matthaeus (2021) for a
review). In the theory of energetic particle transport all scales of magnetic turbulence can be significant. For perpendicular transport
the large scales of the energy range are particularly important. This is due to the relevance of magnetic field line random walk (FLRW)
as discussed in the next paragraph.

The simplest picture of perpendicular transport is that of the \textit{field line random walk limit}. Magnetic field lines in turbulence
are stochastic curves. The properties of those curves are mostly controlled by the large scale structure of turbulence (see, e.g.,
Shalchi \& Kourakis (2007b,c)). Often it is assumed that magnetic field lines behave diffusively. The term \textit{diffusion}, however,
is often causing confusion. Diffusion means in this case a linear increase of the uncertainty to find a magnetic field line at a certain
location of space. In terms of mean square displacements, one has in the diffusive case $\langle ( \Delta x ) \rangle = 2 \kappafl \left| z \right|$
where $\kappafl$ is the field line diffusion coefficient. Here $z$ is the coordinate along the mean magnetic field and $x$ corresponds
to the perpendicular direction, respectively. Non-diffusive field line random walk has been discussed in the literature (see, e.g.,
Zimbardo \& Veltri (1995), Zimbardo et al. (1995), Zimbardo et al. (2000), Zimbardo (2005), and Shalchi \& Kourakis (2007b,c)).
The understanding of the magnetic field line behavior itself is complicated due to the non-linearity of the problem (see, e.g.,
Kadomtsev \& Pogutse (1979), Mattheaus et al. (1995), and Shalchi \& Kourakis (2007b,c)). Within the field line random walk limit,
one assumes that the particles follow magnetic field lines while their motion along the mean magnetic field occurs with constant velocity.
If this was true, the perpendicular diffusion coefficient of the particle is given by
\be
\kappa_{\perp} = \frac{v}{2}\kappafl \quad\quad \textnormal{(FLRW Limit)}
\label{FLRWlimit}
\ee
where $v$ is the particle speed. This means that perpendicular transport is entirely controlled by the behavior of stochastic magnetic
field lines. Sometimes this type of transport is called \textit{first diffusion} (see, e.g., Matthaeus et al. (2003)). However, in reality,
the particle motion is not unperturbed in the parallel direction. In laboratory plasmas, for instance, Coulomb collisions lead to parallel
diffusion (see, e.g., Rechester \& Rosenbluth (1978)). In the context of space and astrophysical plasmas collisions are absent but instead
there is pitch-angle scattering due to the interaction between particles and turbulent magnetic fields (see, e.g., Shalchi (2009) for a review).
If pitch-angle scattering occurs over an extended period of time, the particle motion becomes diffusive in the parallel direction. As a
consequence, perpendicular transport is suppressed to a sub-diffusive level. This type of transport is usually referred to as \textit{compound
sub-diffusion} and can be described by using analytical theory (see Webb et al. (2006) for the most comprehensive description of this type of
transport). 

In test-particle simulations one can nicely observe how particle transport is indeed sub-diffusive in some cases but normal diffusion is restored
in other cases (see Qin et al. (2002a,b)). Therefore, the question arises what triggers the recovery of Markovian diffusion? In the case of Coulomb
collisions the particles can get scattered off the original magnetic field line they were tied to. The particle then \textit{jumps} to a neighbor
field line. If magnetic field lines separate, the particle motion becomes more complex leading to normal diffusion instead of compound sub-diffusion.
This \textit{picture} of perpendicular transport forms the basis of the famous Rechester \& Rosenbluth (1978) paper. In the context of space and
astrophysics, collisions are absent as described already above. Furthermore, magnetic field lines do not separate exponentially as assumed by
Rechester \& Rosenbluth (1978). However, as described in Shalchi (2019a) and Shalchi (2020a,b), Coulomb collisions and exponentially separating
field lines are not needed in order to restore diffusion. Pitch-angle scattering itself can scatter particles away from the original field
line they were tied to. As long as field lines separate, particle diffusion across the mean field will be restored. Everything what is needed in order
to recover diffusion is a combination of pitch-angle scattering and transverse complexity of the turbulence. The latter effect is directly related
to field line separation (see, e.g., Shalchi (2019b)). In this scenario, sometimes called \textit{second diffusion} (see again Matthaeus et al. (2003)),
one has (see Shalchi (2019a)) $\kappa_{\perp} / \kappa_{\parallel} = \langle ( \Delta x )^2 \rangle / \langle ( \Delta z )^2 \rangle \approx \ell_{\perp}^2 / L_K^2$
where we have used the perpendicular correlation scale of the turbulence $\ell_{\perp}$ as well as the \textit{Kolmogorov-Lyapunov length} $L_K$.
The latter scale indicates the distance one needs to travel along the mean field in order to observe significant transverse structure of the turbulence.
Since particles follow field lines until transverse complexity becomes important, one can estimate $\ell_{\perp}^2 \approx \kappafl L_K$. Using this
above to eliminate $L_K$ leads to
\be
\frac{\kappa_{\perp}}{\kappa_{\parallel}} = \frac{\kappafl^2}{\ell_{\perp}^2} \quad\quad \textnormal{(CLRR Limit)}
\label{CLRR}
\ee
which can be understood as collisionless version of Rechester \& Rosenbluth (1978) and, therefore, we refer to the latter limit as the
\textit{collisionless Rechester \& Rosenbluth (CLRR) limit}. According to Eq. (\ref{CLRR}), the ratio of perpendicular and parallel diffusion
coefficients is constant and, in most cases, small.

A very critical parameter in the theory of three-dimensional turbulence is the so-called \textit{Kubo number} which was introduced by
Kubo (1963) and is defined via
\be
K = \frac{\ell_{\parallel}}{\ell_{\perp}} \frac{\delta B_x}{B_0} \quad\quad \textnormal{(Kubo Number)}
\label{defineKuboNumber}
\ee
where we have used the turbulence correlation scales in parallel and perpendicular directions, the $x$-component of the turbulent
magnetic field $\delta B_x$, and the mean field $B_0$. One expects to find the field line random walk limit as given by Eq. (\ref{FLRWlimit})
for very long parallel mean free paths and CLRR diffusion as given by Eq. (\ref{CLRR}) for short parallel mean free paths. Analytical theories,
on the other hand, provide in the limit of short parallel mean free paths and large Kubo numbers the result
\be
\frac{\kappa_{\perp}}{\kappa_{\parallel}} = \frac{\delta B_x^2}{B_0^2} \quad\quad \textnormal{(Fluid Limit)}
\label{Fluid}
\ee
as, for instance, derived in Shalchi et al. (2004), Zank et al. (2004), Shalchi et al. (2010), and Shalchi (2015). In this case particles
follow ballistic magnetic field lines while their parallel motion is diffusive (see Shalchi (2019a) and Shalchi (2020a,b)). A remaining
puzzle in the theory of perpendicular transport is how analytical theories can be altered so that they provide Eq. (\ref{CLRR}) and
explain how CLRR and fluid limits are related to each other in the large Kubo number regime. Furthermore, heuristic arguments cannot be very
accurate and in some cases it is unclear how they need to be used in order to construct a perpendicular diffusion coefficient. This is in
particular the case for two-component turbulence.

In reality, perpendicular transport is a combination of first and second diffusion. If pitch-angle scattering is weak, corresponding
to a very long parallel mean free path, perpendicular transport is mostly controlled by the random walk of magnetic field lines and
one finds $\kappa_{\perp} \approx \kappafl v /2$. If pitch-angle scattering is very strong, equivalent to very short parallel mean free
paths, the perpendicular diffusion coefficient is close to $\kappa_{\perp} / \kappa_{\parallel} = const$. Qualitatively, this behavior of
perpendicular transport is universal meaning that it does not depend on the detailed properties of magnetic turbulence (see Hussein et al. (2015)).
However, in turbulence without transverse structure, usually called \textit{slab turbulence}, second diffusion cannot be established due
to the lack of field line separation. In this case compound sub-diffusion is the final state of the transport. It is important to note
that in turbulence with transverse complexity, diffusion will eventually be restored. However, a long time can pass before the
transverse structure of the turbulence starts to influence the particle motion in the perpendicular direction. Therefore, after the
initial ballistic regime, one often observes sub-diffusive behavior for a long time before diffusion is finally restored. The
sequence \textit{ballistic transport $\rightarrow$ sub-diffusion $\rightarrow$ normal diffusion}, is in particular prevalent in
turbulence with small Kubo numbers (see Shalchi (2019a) and Shalchi (2020a,b)).

Although our qualitative understanding of perpendicular transport is complete, one still needs an analytical theory describing the
physics outlined above. Such a theory, in order to be reliable, needs to be in good agreement with test-particle simulations for
a variety of turbulence configurations. It is clear that due to the complexity of the problem, simple concepts such as quasi-linear
theory or hard-sphere scattering models will fail completely in the general case. Major progress has been achieved by Mattheaus et al. (2003),
where after employing a set of approximations, a non-linear integral equation has been derived. The latter theory provided some
promising results in particular in the context of solar wind observations (see, e.g., Bieber et al. (2004)). However, the aforementioned
theory cannot be the final solution to the problem. First of all the theory does not provide a vanishing diffusion coefficient for slab
turbulence as observed in simulations (see again Qin et al. (2002a)). Furthermore, the theory does not provide the correct field line
random walk limit if pitch-angle scattering is suppressed. In other words, the Mattheaus et al. (2003) theory disagrees with the
Mattheaus et al. (1995) theory of field line random walk. In order to address these two problems, the unified non-linear transport (UNLT)
theory has been developed in Shalchi (2010). The theory provides $\kappa_{\perp} = 0$ for slab turbulence and it contains the Mattheaus et al. (1995)
theory as special limit. Therefore, it is a unified transport theory for magnetic field lines and energetic particles. A time-dependent
version of UNLT theory was developed in Shalchi (2017) and Lasuik \& Shalchi (2017). The latter theory was able to provide a full
time-dependent description of the transport including the initial ballistic regime, compound sub-diffusion, as well as the recovery
of diffusion due to transverse complexity. In fact it was the first theory which was able to explain why and how diffusion is restored.
However, all the aforementioned systematic theories have one problem in common. In the strong scattering regime where the parallel mean
free path is short, and for turbulence with large Kubo numbers (this includes the case of two-component turbulence), the theories do not
agree very well with simulations. In fact they provide a perpendicular diffusion coefficient which is roughly a factor $3$ too large
(see, for instance, Figs. \ref{SimComp1}-\ref{SimComp6} of the current paper). In order to balance out this discrepancy, a correction
factor $a^2 \approx 1/3$ was included into previous transport theories. It is the purpose of the current article to develop a theory
which does not require this type of correction factor but has all the advantages of UNLT theories. 

The reminder of the present article is organized as follows. In Section 2 we provide an overview over previously developed analytical
descriptions of perpendicular transport. This include a discussion of the field line random walk limit, diffusive UNLT theory, and
its time-dependent version. In Section 3 a new theory is systematically developed and in the following Sections 4-6 special cases
and limits are considered. Section 7 focuses on a comparison with previously performed test-particle simulations for the well-known
and accepted two-component turbulence model. In Section 8 we summarize and conclude.

\section{Previously Developed Descriptions of Perpendicular Transport}

In order to achieve a further improvement of the analytical theory of perpendicular transport, it is essential to briefly review
existing theories.

\subsection{The FLRW Limit}

The simplest picture of perpendicular transport is that of the \textit{field line random walk limit}. In this case particles are tied
to magnetic field lines while their parallel motion occurs unperturbed (with constant velocity). Since the random walk of magnetic field
lines is often assumed to behave diffusively, described by the field line diffusion coefficient $\kappafl$, the particles move diffusively
in the perpendicular direction. The corresponding particle diffusion coefficient is then given by Eq. (\ref{FLRWlimit}). To understand
the field line random walk itself is complicated due to the non-linearity of the problem. Only in slab turbulence, corresponding to
turbulence without any transverse complexity, the theory of field line random walk is exact (see Shalchi (2020a) for a detailed
explanation). However, based on Corrsin's independence hypothesis (see Corrsin (1959)), sometimes called \textit{random phase
approximation}, Kadomtsev \& Pogutse (1979) as well as Matthaeus et al. (1995) developed a theory for the field line diffusion
coefficient. The latter theory provides the following non-linear integral equation
\be
\kappafl = \frac{1}{B_0^2} \int d^3 k \; P_{xx} \big( \vec{k} \big)
\frac{\kappafl k_{\perp}^2}{k_{\parallel}^2 + (\kappafl k_{\perp}^2)^2}.
\label{bill95}
\ee
This equation contains two asymptotic limits depending on what the Kubo number is. The first asymptotic limit is obtained for small Kubo
numbers for which Eq. (\ref{bill95}) turns into the quasi-linear limit
\be
\kappafl = L_{\parallel} \frac{\delta B_x^2}{B_0^2} \quad\quad \textnormal{(for $K \ll 1$)}
\label{kappaflLpara}
\ee
where we have used the parallel integral scale $L_{\parallel}$ of the turbulence. For large Kubo numbers, on the other hand, one obtains
\be
\kappafl = L_U \frac{\delta B_x}{B_0} \quad\quad \textnormal{(for $K \gg 1$)}
\label{kappaflLU}
\ee
where we have used the ultra-scale $L_U$. By combining Eqs. (\ref{FLRWlimit}) and (\ref{bill95}) one can easily derive an integral equation
for the perpendicular diffusion coefficient of the charged particles namely
\be
\kappa_{\perp} = \frac{v^2}{3 B_0^2} \int d^3 k \; \frac{P_{xx} \big( \vec{k} \big)}{ v^2 k_{\parallel}^2 / (3 \kappa_{\perp} k_{\perp}^2) + (4/3) \kappa_{\perp} k_{\perp}^2}.
\label{FLRWlimiteq}
\ee
In the following we refer to Eq. (\ref{FLRWlimiteq}) either as the field line random walk limit or the weak (pitch-angle) scattering limit.
This corresponds to the formal limit where the parallel mean free path is $\lambda_{\parallel} \rightarrow \infty$. 

\subsection{Diffusive UNLT Theory}

The basic equation describing the particle motion in a magnetic system is the Newton-Lorentz equation. However, a significant simplification can be achieved by
using \textit{guiding center coordinates} (see, e.g., Schlickeiser (2002)) so that the equation of motion for perpendicular transport becomes
\be
V_x (t) = v_z (t) \frac{\delta B_x \big[ \vec{x} \big( t \big) \big]}{B_0}
\label{eqofmotion}
\ee
where $V_x (t)$ is the perpendicular velocity of the guiding center. A very similar equation can be obtained for $V_y (t)$ but this
is not necessary due to the usual assumption of axi-symmetry. In the following we perform some steps in order to rewrite Eq. (\ref{eqofmotion}).
These steps are standard in transport theory (see, e.g., Shalchi (2020a) for a review). First we replace the turbulent magnetic field in
Eq. (\ref{eqofmotion}) by a usual Fourier representation
\be
\delta B_x \big( \vec{x} \big) = \int d^3 k \; \delta B_x (\vec{k}) e^{i \vec{k} \cdot \vec{x}}.
\label{Fourier}
\ee
Using this in the equation of motion (\ref{eqofmotion}) yields
\be
V_x (t) = \frac{1}{B_0} \int d^3 k \; \delta B_x (\vec{k}) v_z (t) e^{i \vec{k} \cdot \vec{x}}.
\label{eqofmotion2}
\ee
From this one can derive the \textit{velocity auto-correlation function} as
\bdm
\left< V_x (t) V_x (0) \right> & = & \frac{1}{B_0^2} \int d^3 k \int d^3 k^{\prime} \; \left< \delta B_x (\vec{k}) \delta B_x^* (\vec{k}^{\prime}) \right. \nonumber\\
& \times & \left. v_z (t) v_z (0) e^{i \vec{k} \cdot \vec{x} (t) - i \vec{k}^{\prime} \cdot \vec{x} (0)} \right>.
\label{VCF}
\edm
Therein one can easily see the emergence of higher order correlation functions. In order to simplify this, we employ \textit{Corrsin's independence hypothesis}
(see Corrsin (1959)) allowing us to write
\bdm
\left< V_x (t) V_x (0) \right> & = & \frac{1}{B_0^2} \int d^3 k \int d^3 k^{\prime} \; \left< \delta B_x (\vec{k}) \delta B_x^* (\vec{k}^{\prime}) \right> \nonumber\\
& \times & \left< v_z (t) v_z (0) e^{i \vec{k} \cdot \vec{x} (t) - i \vec{k}^{\prime} \cdot \vec{x} (0)} \right>.
\label{VCFCorrsin}
\edm
To continue we assume homogeneous turbulence to find
\bdm
\left< V_x (t) V_x (0) \right> & = & \frac{1}{B_0^2} \int d^3 k \; P_{xx} (\vec{k}) \nonumber\\
& \times & \left< v_z (t) v_z (0) e^{i \vec{k} \cdot \left[ \vec{x} (t) - \vec{x} (0) \right]} \right>
\label{VCF2}
\edm
where we have used the spectral tensor $P_{xx} (\vec{k})$. As the next step one can employ the {\it TGK (Taylor-Green-Kubo) formulation}
(see Taylor (1922), Green (1952), and Kubo (1957))
\be
d_{\perp} (t) = d_{\perp} (0) + \Re \int_{0}^{t} d t' \; \big< V_x (t') V_x (0) \big>
\label{tgk}
\ee
where we have used the running perpendicular diffusion coefficient defined via
\be
d_{\perp} (t) = \frac{1}{2} \frac{d}{d t} \big< \left( \Delta x \right)^2 \big>.
\label{definerunning}
\ee
In Eq. (\ref{tgk}) we have allowed for a non-vanishing initial diffusion coefficient $d_{\perp} (0)$. However, perpendicular transport
is initially ballistic and, therefore, $d_{\perp} (0) = 0$ in our case. In the limit $t \rightarrow \infty$ we find the usual
time-independent diffusion coefficient
\be
\kappa_{\perp} = \lim_{t \rightarrow \infty} d_{\perp} (t) = \Re \int_{0}^{\infty} d t \; \left< V_x (t) V_x (0) \right>.
\label{kappaperpdperp}
\ee
After combining Eqs. (\ref{VCF2}) and (\ref{kappaperpdperp}), we obtain
\be
\kappa_{\perp} = \frac{v^2}{B_0^2} \int d^3 k \; P_{xx} (\vec{k}) T \big( \vec{k} \big)
\label{withdperp0}
\ee
where we have used
\be
T \big( \vec{k} \big) = \frac{1}{v^2} \int_{0}^{\infty} d t \; \left< v_z (t) v_z (0) e^{i \vec{k} \cdot \left[ \vec{x} (t) - \vec{x} (0) \right]} \right>.
\label{definexi}
\ee
The key problem is to determine the function $T (\vec{k})$. Important here is to note that velocities and positions are highly correlated.
Therefore, a pitch-angle dependent transport equation is needed. In Shalchi (2010) the following Fokker-Planck equation was used in order
to determine the analytical form of $T (\vec{k})$
\be
\frac{\partial f}{\partial t} + v \mu \frac{\partial f}{\partial z}
= \frac{\partial}{\partial \mu} \left[ D_{\mu\mu} \frac{\partial f}{\partial \mu} \right]
+ D_{\perp} \left[ \frac{\partial^2 f}{\partial x^2} + \frac{\partial^2 f}{\partial y^2} \right].
\label{FPeq}
\ee
The latter equation and more general forms, are discussed in Skilling (1975), Schlickeiser (2002), and Zank (2014). In Eq. (\ref{FPeq})
we have used the particle distribution function $f$, time $t$, the particle speed $v$, the pitch-angle cosine $\mu$, and the position in three-dimensional
space $(x,y,z)$. Furthermore, the equation contains the pitch-angle Fokker-Planck coefficient $D_{\mu\mu}$ as well as the Fokker-Planck coefficient
of perpendicular diffusion $D_{\perp}$. The quantity $T (\vec{k})$ defined via Eq. (\ref{definexi}) can be derived form Eq. (\ref{FPeq}) as 
shown in Shalchi (2010). After some lengthy algebra on obtains the following form
\be
T \big( \vec{k} \big) = \frac{1}{3} \frac{1}{v / \lambda_{\parallel} + (v k_{\parallel})^2 /(3 \kappa_{\perp} k_{\perp}^2) + (4/3) \kappa_{\perp} k_{\perp}^2}.
\label{UNLTxi}
\ee
Combining this with Eq. (\ref{withdperp0}) leads to the following integral equation
\be
\kappa_{\perp} = \frac{a^2 v^2}{3 B_0^2} \int d^3 k \;
\frac{P_{xx} (\vec{k})}{v / \lambda_{\parallel} + v^2 k_{\parallel}^2/(3 \kappa_{\perp} k_{\perp}^2) + (4/3) \kappa_{\perp} k_{\perp}^2}
\label{UNLT}
\ee
where we have introduced the correction factor $a^2$ to balance out inaccuracies of the used approximations. It will be the aim of the current
paper to develop an analytical theory which does not require this type of factor.

One can easily see that in the asymptotic limit $\lambda_{\parallel} \rightarrow \infty$, Eq. (\ref{UNLT}) reduces indeed to the FLRW limit
given by Eq. (\ref{FLRWlimiteq}). Furthermore, for slab turbulence, we have $k_{\perp} \rightarrow 0$, and Eq. (\ref{UNLT}) provides $\kappa_{\perp} = 0$
as solution as required. Compared to other theories such as the \textit{Non-Linear Guiding Center (NLGC) theory} of Matthaeus et al. (2003), the main
advantage of Eq. (\ref{UNLT}) is that it provides the correct result for slab turbulence but it also works for three-dimensional turbulence with small Kubo
numbers (see Shalchi (2020a)). For two-dimensional turbulence and three-dimensional turbulence with large Kubo numbers, on the other hand,
Eq. (\ref{UNLT}) and NLGC theory are almost identical (see next subsection).

\subsection{UNLT and NLGC Theories for Two-Dimensional Turbulence}

Solar wind fluctuations contain a subpopulation with wave vector orientations nearly perpendicular to the mean magnetic field (Matthaeus et al. (1990)).
As also emphasized in Matthaeus et al. (1990), the corresponding density fluctuations are small corresponding to the nearly incompressible case.
This type of turbulence is usually called two-dimensional (2D) turbulence. Two-dimensional turbulence is also familiar in laboratory
plasma studies such as \textit{tokamaks} (see, e.g., Robinson \& Rusbridge (1971) as well as Zweben et al. (1979)). The theoretical framework for this
type of turbulence is provided by \textit{reduced magnetohydrodynamics} (see Zank \& Matthaeus (1992)). For two-dimensional magnetic fluctuations, the
corresponding spectral tensor is given by
\be
P_{nm} (\vec{k}) = g^{2D} (k_{\perp}) \frac{\delta (k_{\parallel})}{k_{\perp}} \left[ \delta_{nm} - \frac{k_n k_m}{k_{\perp}^2} \right]
\label{Pnm2D}
\ee
where we have used $n,m=x,y$. All other tensor components are assumed to be zero. This is a consequence of considering nearly incompressible
turbulence where $\delta B_z \approx 0$. The function $g^{2D} (k_{\perp})$ corresponds to the two-dimensional spectral function. The analytical
form of the spectrum will be discussed later in this paper. With Eq. (\ref{Pnm2D}) diffusive UNLT theory, represented by Eq. (\ref{UNLT}), becomes
\be
\kappa_{\perp} = \pi \frac{a^2 v^2}{3 B_0^2} \int d^3 k \; g^{2D} \big( k_{\perp} \big) \frac{1}{v / \lambda_{\parallel} + (4/3) \kappa_{\perp} k_{\perp}^2}.
\label{NLGC}
\ee
It needs to be emphasized that for this very special case, the latter equation agrees with the  NLGC theory of Matthaeus et al. (2003) except for
the factor $4/3$ in the denominator. Although this factor needs to be there in order to find agreement with the field line random walk limit
given by Eq. (\ref{FLRWlimiteq}), this is a minor difference.

\subsection{Time-Dependent UNLT Theory}

A further improvement of the theory of perpendicular transport was achieved due to the development of time-dependent UNLT theory 
(see Shalchi (2017) and Lasuik \& Shalchi (2017)). This approach is discussed in the following. As before we use Eq. (\ref{eqofmotion})
as starting point and as in diffusive UNLT theory we employ the same steps until we arrive at Eq. (\ref{VCF}). The fundamental
approximation leading to time-dependent UNLT theory is
\bdm
& & \big< v_z (t) v_z (0) \delta B_x \big( \vec{k} \big) \delta B_x^* \big( \vec{k}' \big)
e^{i z(t) k_{\parallel} + i \vec{x}_{\perp} (t) \cdot \vec{k}_{\perp}} \big> \nonumber\\
& \approx & a^2 \big< v_z (t) v_z (0) e^{i k_z z(t)} \big>
\big< \delta B_x \big( \vec{k} \big) \delta B_x^* \big( \vec{k}' \big) \big>
\big< e^{i \vec{x}_{\perp} (t) \cdot \vec{k}_{\perp}} \big>.
\label{centralUNLT}
\edm
This means that we group together magnetic fields, all particle properties associated with their parallel motion, and all particle
properties associated with their perpendicular motion. In Eq. (\ref{centralUNLT}) we have used again the parameter $a^2$ to balance out inaccuracies
arising from the used approximations. Using Eq. (\ref{centralUNLT}) in (\ref{VCF}) allows us to write the velocity auto-correlation function as
\be
\big< V_x (t) V_x (0) \big> = \frac{a^2}{B_0^2} \int d^3 k \; P_{xx} \big( \vec{k} ,t \big)
\xi \big( k_{\parallel}, t \big) \big< e^{i \vec{x}_{\perp} \cdot \vec{k}_{\perp}} \big>
\label{centraleq}
\ee
where we have used the parallel correlation function
\be
\xi \big( k_{\parallel}, t \big)
= \big< v_z \left( t \right) v_z \left( 0 \right) e^{i z \left( t \right) k_{\parallel}} \big>.
\label{definexiparallel}
\ee
For the perpendicular characteristic function in Eq. (\ref{centraleq}) we employ
\be
\big< e^{i \vec{x}_{\perp} \cdot \vec{k}_{\perp}} \big> = e^{- \frac{1}{2} \langle (\Delta x)^2 \rangle k_{\perp}^2}.
\label{charinUNLT}
\ee
The latter form corresponds to a Gaussian function with zero mean. In Lasuik \& Shalchi (2018) alternative distributions, such as
\textit{kappa distributions}, were considered but it was demonstrated that the assumed statistics has only a very small influence
on the resulting perpendicular diffusion coefficient. If we combine Eqs. (\ref{centraleq}) and (\ref{charinUNLT}), we find for the
velocity correlation function
\be
\langle V_x (t) V_x (0) \rangle = \frac{a^2}{B_0^2} \int d^3 k \; P_{xx} \big( \vec{k}, t \big)
\xi \big( k_{\parallel}, t \big) e^{- \frac{1}{2} \langle ( \Delta x )^2 \rangle k_{\perp}^2}.
\label{centraleq2}
\ee
From Eqs. (\ref{tgk}) and (\ref{definerunning}) it follows that
\be
\frac{d^2}{d t^2} \langle (\Delta x)^2 \rangle = 2 \langle V_x (t) V_x (0) \rangle
\label{relateMSDandVCF}
\ee
and, therefore, we can write Eq. (\ref{centraleq2}) as
\be
\frac{d^2}{d t^2} \langle (\Delta x)^2 \rangle = \frac{2 a^2}{B_0^2} \int d^3 k \; P_{xx} \big( \vec{k}, t \big)
\xi \big( k_{\parallel}, t \big) e^{- \frac{1}{2} \langle (\Delta x)^2 \rangle k_{\perp}^2}.
\label{Generalsigmap}
\ee
The solution of this integro-differential equation is the mean square displacement $\langle ( \Delta x )^2 \rangle$ of the energetic
particle. How this solution looks like depends on the spectral tensor $P_{xx} (\vec{k},t)$, describing magnetic turbulence, and the parallel
correlation function $\xi (k_{\parallel},t)$. It was shown before (see Shalchi (2020a) for a systematic explanation) that
\be
\xi \left( k_{\parallel}, t \right) = \frac{v^2}{3} \frac{1}{\omega_+ - \omega_-}
\left[ \omega_+ e^{\omega_+ t} - \omega_- e^{\omega_- t} \right]
\label{formforWoldapp}
\ee
with the parameters $\omega_{\pm}$ given by
\be
\omega_{\pm} = - \frac{v}{2 \lambda_{\parallel}} \pm \sqrt{\left( \frac{v}{2 \lambda_{\parallel}} \right)^2 - \frac{1}{3} \Big( v k_{\parallel} \Big)^2}
\label{omegapm}
\ee
which requires knowledge of the parallel mean free path $\lambda_{\parallel}$. Note, this result was obtained by solving Eq. (\ref{FPeq}) without
perpendicular diffusion term. The parallel correlation function given by Eqs. (\ref{formforWoldapp}) and (\ref{omegapm}) describes parallel transport
as a process which is initially ballistic and at later times diffusive. Diffusive UNLT theory, represented by Eq. (\ref{UNLT}), can be derived by
combining Eq. (\ref{Generalsigmap}) with a diffusion approximation of the form $\langle (\Delta x)^2 \rangle = 2 \kappa_{\perp} t$ and time-integrating.
From this one can derive
\be
\kappa_{\perp} = \frac{a^2 v^2}{3 B_0^2} \int d^3 k \;
\frac{P_{xx} (\vec{k})}{v / \lambda_{\parallel} + v^2 k_{\parallel}^2/(3 \kappa_{\perp} k_{\perp}^2) + \kappa_{\perp} k_{\perp}^2}.
\label{difffromtimeUNLT}
\ee
This result is almost identical compared to Eq. (\ref{UNLT}). The only difference is the factor $4/3$ in the denominator of Eq. (\ref{UNLT}).
The latter equation is more accurate because the factor $4/3$ is needed in order to find the correct field line random walk limit as given by
Eq. (\ref{FLRWlimiteq}). However, this factor does not change the perpendicular diffusion coefficient significantly. The reason why this
factor is not part of time-dependent UNLT theory is the subspace approximation used to derive Eq. (\ref{formforWoldapp}).

\section{The Field Line - Particle Decorrelation Theory}

Diffusive and time-dependent UNLT theories have been successful in describing perpendicular transport for a wide range of parameters and turbulence
models (see Shalchi (2020a) for a review). The aforementioned theories contain previously developed approaches as special limits. For instance,
the FLRW limit (see Eq. (\ref{FLRWlimiteq}) of the current paper) is obtained by considering $\lambda_{\parallel} \rightarrow \infty$. Compound
sub-diffusion for slab turbulence (see, e.g., Webb et al. (2006)) is automatically obtained for this specific turbulence model. For two-dimensional
turbulence diffusive UNLT theory agrees with the Matthaeus et al. (2003) theory (except for the factor $4/3$ discussed above). Even a collisionless
version of Rechester \& Rosenbluth (1978) can be obtained if the assumption of strong pitch-angle scattering is combined with small Kubo number
turbulence (see Shalchi (2015)).

However, UNLT theory deviates from test-particle simulations in some cases. This is in particular the case if pitch-angle scattering is strong
and if turbulence with larger Kubo numbers is considered. This numerical disagreement (roughly a factor $3$) was previously balanced out by
incorporating a correction factor $a^2 \approx 1/3$ as originally suggested by Matthaeus et al. (2003). It was finally explained in Shalchi (2019a)
why this factor is needed and why previously developed theories fail in the aforementioned case. In the context of diffusive UNLT theory the reason
is that Eq. (\ref{FPeq}) is not valid in the general case. This problem is often overlooked in the literature. Eq. (\ref{FPeq}) is based on the
assumption that perpendicular transport becomes diffusive instantaneously. After some time has passed, corresponding to the pitch-angle
isotropization time scale, parallel transport becomes diffusive as well. Therefore, Eq. (\ref{FPeq}) is valid as long as perpendicular transport
becomes diffusive before parallel transport. However, this is not always true. In particular for short parallel mean free paths, parallel transport
becomes diffusive before perpendicular transport. If particles follow magnetic field lines while parallel transport is diffusive, perpendicular
transport is suppressed to a sub-diffusive level. Perpendicular transport is then restored later as soon as the transverse complexity of the
turbulence becomes significant. This scenario is not described by Eq. (\ref{FPeq}). Therefore, it is important that non-diffusive transport
equations are explored (see, e.g., Webb et al. (2006), Zimbardo et al. (2017), and Strauss \& Effenberger (2017)).

In the context of time-dependent UNLT theory, the problem is that approximation (\ref{centralUNLT}) does not work in the general case. In reality
particle trajectories and particle velocities are highly correlated. If this correlation is neglected, resulting theories deviate from test-particle
simulations as demonstrated in detail in Qin \& Shalchi (2016). The assumption that trajectories and velocities are uncorrelated is also the reason
why the NLGC theory of Matthaeus et al. (2003) needs the correction factor $a^2$ and why it does not work for slab and small Kubo number turbulence.

\subsection{Generalized Compound Sub-Diffusion}

A detailed analysis of compound sub-diffusion was presented in Webb et al. (2006) who have developed a description for this type of transport
based on the so-called \textit{Chapman-Kolmogorov equation} (see, e.g., Gardiner (1985))
\be
f_{\perp} (x,y;t) =  \int_{-\infty}^{+\infty} d z \; f_{\scriptscriptstyle F\!L} (x,y;z) f_{\parallel} (z;t)
\label{chapman}
\ee
where the particle distribution in the perpendicular direction $f_{\perp} (x,y;t)$ is given as convolution integral of the
parallel distribution function $f_{\parallel} (z;t)$ and the field line distribution function $f_{\scriptscriptstyle F\!L} (x,y;z)$.
The parallel distribution function describes the probability to find the particle at the (parallel) position $z$ at a certain
time $t$. In the past (see again Webb et al. (2006)) a Gaussian distribution with vanishing mean has been used to model the parallel
motion and the function $f_{\parallel} (z;t)$. However, parallel transport is a process which is initially ballistic and becomes
diffusive (Gaussian) at later times. Therefore, to model the function $f_{\parallel} (z;t)$ is not trivial. More details about the
analytical form of the parallel distribution and its Fourier transform can be found below. Eq. (\ref{chapman}) corresponds to the
assumption that particles follow magnetic field lines and that the particle distribution in the perpendicular direction depends
only on parallel transport and the field line random walk.

From Eq. (\ref{chapman}) we can consider the second moment to find (see, e.g., Shalchi \& Kourakis (2007a))
\be
\big< \left( \Delta x \right)^2 \big>_P \big( t \big)
= \int_{-\infty}^{+\infty} d z \; \big< \left( \Delta x \right)^2 \big>_{\scriptscriptstyle F\!L} \big( z \big) f_{\parallel} (z;t).
\label{chapmanMSD}
\ee
As in Eq. (\ref{chapman}) we have used the parallel distribution function of the particles therein. The quantity $\langle \left( \Delta x \right)^2 \rangle_P$
corresponds to the perpendicular mean square displacement of the energetic particles and $\langle \left( \Delta x \right)^2 \rangle_{\scriptscriptstyle F\!L}$
to the mean square displacement of the magnetic field lines. The parallel distribution function in Eq. (\ref{chapmanMSD}) can be replaced by the Fourier
representation
\be
f_{\parallel} \big( z; t \big) = \int_{-\infty}^{+\infty} d k_{\parallel}^{\prime} \; F \big( k_{\parallel}^{\prime}, t \big) e^{i k_{\parallel}^{\prime} z}
\label{fparallelFourier}
\ee
where we have used the notation $k_{\parallel}^{\prime}$ to distinguish the latter parameter from the wave number used later in this article.
Using Eq. (\ref{fparallelFourier}) in Eq. (\ref{chapmanMSD}) yields
\bdm
& & \big< \left( \Delta x \right)^2 \big>_P \big( t \big) \nonumber\\
& = & \int_{-\infty}^{+\infty} d z \; \big< \left( \Delta x \right)^2 \big>_{\scriptscriptstyle F\!L} (z)
\int_{-\infty}^{+\infty} d k_{\parallel}^{\prime} \; F \big( k_{\parallel}^{\prime}, t \big) e^{i k_{\parallel}^{\prime} z} \nonumber\\
& = & \int_{-\infty}^{+\infty} d k_{\parallel}^{\prime} \; F \big( k_{\parallel}^{\prime}, t \big)
\int_{-\infty}^{+\infty} d z \; \big< \left( \Delta x \right)^2 \big>_{\scriptscriptstyle F\!L} e^{i k_{\parallel}^{\prime} z}
\edm
where, in the last line, we just changed the order of the integrals. We can further rewrite this via
\bdm
\big< \left( \Delta x \right)^2 \big>_P
& = & - \int_{-\infty}^{+\infty} d k_{\parallel}^{\prime} \; F \big( k_{\parallel}^{\prime}, t \big) k_{\parallel}^{\prime \; -2} \nonumber\\
& \times & \int_{-\infty}^{+\infty} d z \; \big< \left( \Delta x \right)^2 \big>_{\scriptscriptstyle F\!L} \frac{d^2}{d z^2} e^{i k_{\parallel}^{\prime} z}
\edm
where we have omitted the time-dependence of $\langle ( \Delta x )^2 \rangle_p$. We continue by using integration by parts twice to find
\bdm
\big< \left( \Delta x \right)_{\scriptscriptstyle F\!L}^2 \big>_P
& = & - \int_{-\infty}^{+\infty} d k_{\parallel}^{\prime} \; F \big( k_{\parallel}^{\prime}, t \big) k_{\parallel}^{\prime \; -2} \nonumber\\
& \times & \int_{-\infty}^{+\infty} d z \; \left[ \frac{d^2}{d z^2} \big< \left( \Delta x \right)^2 \big>_{\scriptscriptstyle F\!L} \right]
e^{i k_{\parallel}^{\prime} z}.
\label{anotherstep1}
\edm
Shalchi \& Kourakis (2007b) derived the following second-order differential equation for the field line mean square displacement
\bdm
& & \frac{d^2}{d z^2} \big< \big( \Delta x \big)^2 \big>_{\scriptscriptstyle F\!L} \nonumber\\
& = & \frac{2}{B_0^2} \int d^3 k \; P_{xx} \big( \vec{k} \big) \cos \big( k_{\parallel} z \big)
e^{- \frac{1}{2} \langle \left( \Delta x \right)^2 \rangle_{\scriptscriptstyle F\!L} k_{\perp}^2}.
\label{Kourakis2007}
\edm
The latter equation describes the random walk of magnetic field lines as a process which is initially ballistic and becomes diffusive at larger
distances $z$. In order to simplify Eq. (\ref{Kourakis2007}), we can use a diffusion approximation for the field lines meaning that we replace
$\langle \left( \Delta x \right)^2 \rangle_{\scriptscriptstyle F\!L} = 2 \kappafl \left| z \right|$ on the right-hand-side. If we would additionally
integrate Eq. (\ref{Kourakis2007}) over all $z$, we would obtain Eq. (\ref{bill95}). Using the field line diffusion approximation and the non-linear
field line equation (\ref{Kourakis2007}) in Eq. (\ref{anotherstep1}) yields
\bdm
\left< \left( \Delta x \right)^2 \right>_P
& = & - \frac{2}{B_0^2} \int_{-\infty}^{+\infty} d k_{\parallel}^{\prime} \; F \big( k_{\parallel}^{\prime}, t \big) k_{\parallel}^{\prime \; -2} \nonumber\\
& \times & \int d^3 k \; P_{xx} \big( \vec{k} \big) \nonumber\\
& \times & \int_{-\infty}^{+\infty} d z \; \cos \big( k_{\parallel} z \big) e^{- \kappafl \left| z \right| k_{\perp}^{2} + i k_{\parallel}^{\prime} z}.\nonumber\\
\label{anotherstep2}
\edm
The $z$-integral therein can be solved via
\bdm
& & \int_{-\infty}^{+\infty} d z \; \cos \big( k_{\parallel} z \big) e^{- \kappafl \left| z \right| k_{\perp}^{2} + i k_{\parallel}^{\prime} z} \nonumber\\
& = & \frac{\kappafl k_{\perp}^{2}}{\big( k_{\parallel} + k_{\parallel}^{\prime} \big)^2 + \left( \kappafl k_{\perp}^{2} \right)^2} + \frac{\kappafl k_{\perp}^{2}}{\big( k_{\parallel} - k_{\parallel}^{\prime} \big)^2 + \left( \kappafl k_{\perp}^{2} \right)^2}.\nonumber\\
\label{beforeresonance}
\edm
For convenience we define the \textit{resonance function}
\be
R_{\pm} \big( \vec{k}^{\prime}, k_{\parallel} \big)
:= \frac{\kappafl k_{\perp}^{2}}{\big( k_{\parallel} \pm k_{\parallel}^{\prime} \big)^2 + \left( \kappafl k_{\perp}^{2} \right)^2}.
\label{resonancefunction}
\ee
In the literature this type of resonance function is usually called a \textit{Cauchy distribution}. Note, the resonance function found here
describes the coupling between particles and magnetic field lines. Using Eqs. (\ref{beforeresonance}) and (\ref{resonancefunction}) in
Eq. (\ref{anotherstep2}) allows us to write
\bdm
\big< \left( \Delta x \right)^2 \big>
& = & - \frac{2}{B_0^2} \int_{-\infty}^{+\infty} d k_{\parallel}^{\prime} \; F \big( k_{\parallel}^{\prime}, t \big) k_{\parallel}^{\prime \; -2} \nonumber\\
& \times & \int d^3 k^{\prime} \; P_{xx} \big( \vec{k} \big) \left[ R_{+} \big( \vec{k}, k_{\parallel}^{\prime} \big) + R_{-} \big( \vec{k}, k_{\parallel}^{\prime} \big) \right].\nonumber\\
\label{anotherstep3}
\edm
Note, Eq. (\ref{anotherstep3}) is just a different way of writing down Eq. (\ref{chapmanMSD}) which, in turn, corresponds to the second moment
of Eq. (\ref{chapman}). Furthermore, we have omitted the subscript \textit{P} meaning that the mean square displacement in Eq. (\ref{anotherstep3}) 
is, of course, the perpendicular mean square displacement of the particles.

\subsection{Velocity Correlations and Parallel Distribution Functions}

Because of reasons that will become clearer later on, we want to derive a relation for the perpendicular velocity correlation function which
can be related to mean square displacements via Eq. (\ref{relateMSDandVCF}). However we keep the mean square displacement in our equation but
consider the second time-derivative so that we derive from Eq. (\ref{anotherstep3})
\bdm
\frac{d^2}{d t^2} \Big< ( \Delta x )^2 \Big>
& = & - \frac{2}{B_0^2} \int_{-\infty}^{+\infty} d k_{\parallel}^{\prime} \; \frac{d^2}{d t^2} F \big( k_{\parallel}^{\prime}, t \big) k_{\parallel}^{\prime \; -2} \nonumber\\
& \times & \int d^3 k \; P_{xx} \big( \vec{k} \big) \left[ R_{+} \big( \vec{k}, k_{\parallel}^{\prime} \big) + R_{-} \big( \vec{k}, k_{\parallel}^{\prime} \big) \right].\nonumber\\
\label{anotherstep4}
\edm
We also need to think about the parallel distribution function in Fourier space $F (k_{\parallel}^{\prime},t)$. Parallel transport of energetic particles
is described via the Fokker-Planck equation
\be
\frac{\partial f}{\partial t} + v \mu \frac{\partial f}{\partial z}
= \frac{\partial}{\partial \mu} \left[ D_{\mu\mu} \frac{\partial f}{\partial \mu} \right]
\label{FPeqnoperp}
\ee
which is a special case of Eq. (\ref{FPeq}). As demonstrated for instance in Shalchi (2020a), Eq. (\ref{FPeqnoperp}) describes the transport
as a process which is initially ballistic. If pitch-angle scattering occurs over an extended period of time, one finds a pitch-angle isotropization process
leading to parallel diffusion. To solve Eq. (\ref{FPeqnoperp}) analytically is difficult but a two-dimensional subspace approximation method has been
developed in the past (see again Shalchi (2020a) for more details). The latter method allows one to derive a simple but accurate solution for the
Fourier-transformed parallel distribution function, namely
\be
F \big( k_{\parallel}, t \big) = \frac{1}{2 \pi} \frac{1}{\omega_+ - \omega_-}
\Big[ \omega_+ e^{\omega_- t} - \omega_- e^{\omega_+ t} \Big].
\label{formforW}
\ee
The latter solution depends on two parameters $\omega_{\pm}$ are given by Eq. (\ref{omegapm}). Note this is related to but not identical compared
to Eq. (\ref{formforWoldapp}). The second time-derivative of Eq. (\ref{formforW}) is
\be
\frac{d^2}{d t^2} F \big( k_{\parallel}, t \big)
= \frac{1}{2 \pi} \frac{\omega_+ \omega_-}{\omega_+ - \omega_-} \left[ \omega_- e^{\omega_- t} - \omega_+ e^{\omega_+ t} \right].
\ee
Therein we can use
\be
\omega_{+} \omega_{-} = \frac{1}{3} v^2 k_{\parallel}^{\prime \; 2}
\ee
which follows from Eq. (\ref{omegapm}). Using these two relations as well as Eq. (\ref{formforWoldapp}) in Eq. (\ref{anotherstep4}) gives us
\bdm
\frac{d^2}{d t^2} \Big< ( \Delta x )^2 \Big>
& = & \frac{1}{\pi B_0^2} \int_{-\infty}^{+\infty} d k_{\parallel}^{\prime} \; \xi \big( k_{\parallel}^{\prime}, t \big) \nonumber\\
& \times & \int d^3 k \; P_{xx} \big( \vec{k} \big) \left[ R_{+} \big( \vec{k}, k_{\parallel}^{\prime} \big) + R_{-} \big( \vec{k}, k_{\parallel}^{\prime} \big) \right].\nonumber\\
\label{anotherstep6}
\edm
One can still understand Eq. (\ref{anotherstep6}) as a generalized compound transport model where parallel transport is initially ballistic and becomes
diffusive for later times. The magnetic field lines have similar properties. A slightly different derivation of Eq. (\ref{anotherstep6}), based on velocity
correlation functions, is presented in Appendix A of the current paper. However, so far a very important ingredient is missing. In the approach presented
above, we do not allow the particles to get scattered away from the original magnetic field line they have been tied to. In order to include this
essential effect, we need to incorporate the transverse complexity of the turbulence 

\subsection{Transverse Complexity of the Turbulence}

In order to achieve a complete analytical description of perpendicular transport, we need to include the effect of transverse structure of the
turbulence into Eq. (\ref{anotherstep6}). The simplest way to do this via including the factor
\be
e^{- \frac{1}{2} \langle ( \Delta x )^2 \rangle k_{\perp}^2} \quad\quad \textnormal{(transverse complexity factor)}
\label{transversecomp}
\ee
where $\langle ( \Delta x )^2 \rangle$ is the particle mean square displacement in the perpendicular direction. The factor used
here corresponds to a Gaussian distribution of the particles with vanishing mean. If this distribution is combined with a diffusion approximation
(see below), this factor is equal to the Fourier-transformed solution of a usual diffusion equation. In fact the factor used here can also be derived
from transport equations such as Eq. (\ref{FPeq}). The mathematical details can be found in the literature (see, e.g., Shalchi (2020a)).

With the help of Eq. (\ref{relateMSDandVCF}), we can easily see that Eq. (\ref{anotherstep6}) corresponds to the velocity correlation function.
Including the transverse complexity factor means that the velocity correlation function decays rapidly as soon as the particles experience the transverse
structure of the turbulence. We finally find the following differential equation
\bdm
\frac{d^2}{d t^2} \Big< ( \Delta x )^2 \Big> & = & \frac{1}{\pi B_0^2} \int_{-\infty}^{+\infty} d k_{\parallel}^{\prime} \; \xi \big( k_{\parallel}^{\prime}, t \big) \nonumber\\
& \times & \int d^3 k \; P_{xx} \big( \vec{k} \big) \left[ R_{+} \big( \vec{k}, k_{\parallel}^{\prime} \big) + R_{-} \big( \vec{k}, k_{\parallel}^{\prime} \big) \right] \nonumber\\
& \times & e^{- \frac{1}{2} \langle ( \Delta x )^2 \rangle k_{\perp}^2}
\label{holyequation}
\edm
where the parallel correlation function $\xi ( k_{\parallel}^{\prime}, t )$ is given by Eq. (\ref{formforWoldapp}), the resonance functions
$R_{\pm} ( \vec{k}, k_{\parallel}^{\prime} )$ are given by Eq. (\ref{resonancefunction}), and $P_{xx} (\vec{k})$ is the $xx$-component
of the spectral tensor as usual. Eq. (\ref{holyequation}) is the central result of this paper. It contains the complete physics necessary to describe
perpendicular transport. The function $\xi ( k_{\parallel}^{\prime}, t )$ describes parallel transport and contains the initial ballistic regime
as well as the diffusive regime. The spectral tensor $P_{xx} ( \vec{k} )$ contains the information about the properties of the magnetic
fluctuations, and the resonance functions $R_{\pm} ( \vec{k}, k_{\parallel}^{\prime} )$ describe the FLRW and the coupling between
field lines and particles. The exponential in Eq. (\ref{holyequation}) describes the effect of transverse complexity leading to the particles
leaving the original magnetic field lines they were tied to. In principle we can solve Eq. (\ref{holyequation}) to obtain the mean square
displacement $\langle ( \Delta x )^2 \rangle$ as a function of time. Compared to time-dependent UNLT theory, however, we need
to evaluate an extra integral. We call the new approach developed here \textit{Field Line - Particle Decorrelation (FLPD) theory}.
Eq. (\ref{holyequation}) corresponds to the time-dependent version of this theory. A diffusive version, where Eq. (\ref{holyequation})
is combined with a diffusion approximation, is discussed in Sect. \ref{DiffApproxAgain}.

\section{Special Limits and Cases}

In the next few paragraphs we consider some limits in order to simplify Eq. (\ref{holyequation}). This will also help the reader to better
understand Eq. (\ref{holyequation}) and the physics it describes.

\subsection{The Limit of Small Kubo Numbers}

\begin{figure}[ht]
\centering
\includegraphics[width=0.50\textwidth]{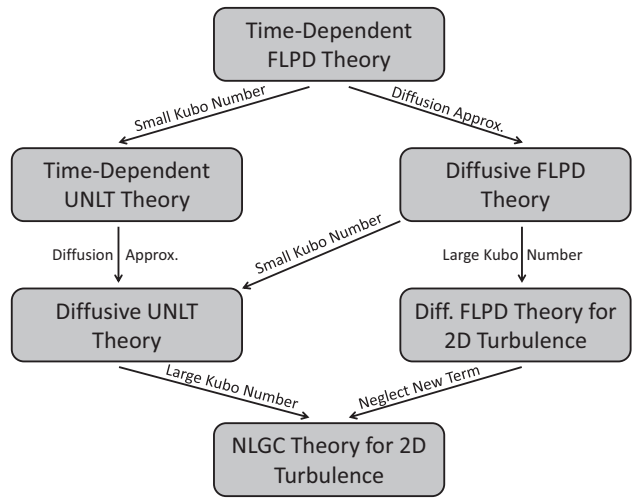}
\caption{This diagram explains how different theories for perpendicular transport are related to each other. Compared are the 
\textit{Field Line - Particle Decorrelation (FLPD) theory} developed in the current paper, the \textit{unified non-linear transport (UNLT) theory},
and the \textit{Non-Linear Guiding Center (NLGC) theory}. FLPD and UNLT theories have time-dependent and diffusive versions.}
\label{TheDiagram}
\end{figure}

Eq. (\ref{holyequation}) depends on the resonance function given by Eq. (\ref{resonancefunction}). If the spectral tensor corresponds to
three-dimensional turbulence with small Kubo numbers, we have by definition $k_{\perp} \rightarrow 0$. Therefore, we can use the relation
\be
\lim_{\alpha \rightarrow 0} \frac{\alpha}{\alpha^2 + x^2} = \pi \delta \left( x \right)
\ee
and our resonance functions (\ref{resonancefunction}) become in the considered limit
\be
R_{\pm} \big( k_{\parallel}^{\prime}, k_{\parallel} \big) = \pi \delta \big( k_{\parallel} \pm k_{\parallel}^{\prime} \big).
\label{smallKuboresonance}
\ee
Using this in Eq. (\ref{holyequation}) yields
\bdm
\frac{d^2}{d t^2} \Big< ( \Delta x )^2 \Big>
& = & \frac{1}{B_0^2} \int_{-\infty}^{+\infty} d k_{\parallel}^{\prime} \; \xi \big( k_{\parallel}^{\prime}, t \big) \nonumber\\
& \times & \int d^3 k \; P_{xx} \big( \vec{k} \big) \left[ \delta \big( k_{\parallel} + k_{\parallel}^{\prime} \big)
+ \delta \big( k_{\parallel} - k_{\parallel}^{\prime} \big) \right] \nonumber\\
& \times & e^{- \frac{1}{2} \langle ( \Delta x )^2 \rangle k_{\perp}^2}.
\edm
Due to the two Dirac deltas, the $k_{\parallel}^{\prime}$-integral can be evaluated to find
\be
\frac{d^2}{d t^2} \Big< ( \Delta x )^2 \Big>
= \frac{2}{B_0^2} \int d^3 k \; P_{xx} \big( \vec{k} \big) \xi \left( k_{\parallel}, t \right) e^{- \frac{1}{2} \langle ( \Delta x )^2 \rangle k_{\perp}^2}
\label{rederivetimeUNLT}
\ee
where we assumed that the spectral tensor is symmetric in $k_{\parallel}$. After comparing this result with Eq. (\ref{Generalsigmap}), we conclude
that for three-dimensional turbulence with small Kubo numbers, our new approach agrees perfectly with time-dependent UNLT theory. This is a very
important result because it was demonstrated in the past that UNLT theory agrees very well with simulations performed for small Kubo number turbulence
without the need of the correction factor $a^2$ (see, e.g., Shalchi (2020a)). For pure slab turbulence, Eq. (\ref{rederivetimeUNLT}) describes
compound sub-diffusion and normal diffusion is never restored. In the next section we shall see that for the opposite case, where the Kubo number
is large, a result is found which is different compared to UNLT theory. The relation between different theories for perpendicular transport
is depicted in Fig. \ref{TheDiagram}.

\subsection{Two-Dimensional Turbulence}

In the previous section we have shown that our new equation for perpendicular transport, namely Eq. (\ref{holyequation}), is identical compared to
time-dependent UNLT theory for small Kubo number turbulence. In the current section we consider the opposite case, namely two-dimensional (2D) turbulence
corresponding to turbulence with large Kubo numbers. In the case of 2D turbulence the corresponding spectral tensor is given by Eq. (\ref{Pnm2D}).
The analytical form of the spectrum $g^{2D} (k_{\perp})$ therein will be discussed later in this paper. With Eq. (\ref{Pnm2D}) our equation for
perpendicular transport (\ref{holyequation}) becomes
\bdm
\frac{d^2}{d t^2} \Big< ( \Delta x )^2 \Big>
& = & \frac{1}{B_0^2} \int_{-\infty}^{+\infty} d k_{\parallel} \; \xi \left( k_{\parallel}, t \right) \nonumber\\
& \times & \int_{0}^{\infty} d k_{\perp} \; g^{2D} (k_{\perp}) \left[ R_{+} \big( \vec{k} \big) + R_{-} \big( \vec{k} \big) \right] \nonumber\\
& \times & e^{- \frac{1}{2} \langle ( \Delta x )^2 \rangle k_{\perp}^2}
\label{timedependentfor2D}
\edm
where now 
\be
R_{\pm} \big( \vec{k} \big) = \frac{\kappafl k_{\perp}^{2}}{k_{\parallel}^2 + \left( \kappafl k_{\perp}^{2} \right)^2}.
\label{2Dresonancefunction}
\ee
It needs to be emphasized that in two-dimensional turbulence there is no parallel wave number. The parameter $k_{\parallel}$ in the resonance
function, given by Eq. (\ref{2Dresonancefunction}), comes from the Fourier transform of the parallel particle distribution function and the
coupling of the particles with the magnetic field lines. Using Eq. (\ref{2Dresonancefunction}) explicitly in Eq. (\ref{timedependentfor2D})
allows us to write
\bdm
\frac{d^2}{d t^2} \Big< ( \Delta x )^2 \Big>
& = & \frac{2}{B_0^2} \int_{-\infty}^{+\infty} d k_{\parallel} \; \xi \left( k_{\parallel}, t \right) \nonumber\\
& \times & \int_{0}^{\infty} d k_{\perp} \; g^{2D} (k_{\perp}) \frac{\kappafl k_{\perp}^{2}}{k_{\parallel}^2 + \left( \kappafl k_{\perp}^{2} \right)^2} \nonumber\\
& \times & e^{- \frac{1}{2} \langle ( \Delta x )^2 \rangle k_{\perp}^2}.
\label{holy2D}
\edm
Note that compared to time-dependent UNLT theory, as given by Eq. (\ref{Generalsigmap}), the new theory provides a very different equation for
perpendicular transport in two-dimensional as well as large Kubo number turbulence. A further simplification of Eq. (\ref{holy2D}) can be
achieved if a diffusion approximation is employed. This in done in the next section.

\section{Re-establishing the Diffusion Approximation}\label{DiffApproxAgain}

In Matthaeus et al. (2003) and Shalchi (2010) a diffusion approximation has been used in order to derive equations much simpler compared
to Eq. (\ref{holyequation}). This type of approximation can also be combined with the new theory. This is done in the following.
Thereafter, we consider again the limits of small and large Kubo numbers, respectively.

\subsection{The Diffusion Approximation for the General Case}

In order to find a simpler approach compared to Eq. (\ref{holyequation}), we can employ a diffusion approximation. Using
$\langle (\Delta x)^2 \rangle = 2 \kappa_{\perp} t$ and integrating Eq. (\ref{holyequation}) over all times yields
\bdm
\kappa_{\perp}
& = & \frac{1}{2 \pi} \frac{1}{B_0^2} \int_{-\infty}^{+\infty} d k_{\parallel}^{\prime} \nonumber\\
& \times & \int d^3 k \; P_{xx} \big( \vec{k} \big) \left[ R_{+} \big( \vec{k}, k_{\parallel}^{\prime} \big) + R_{-} \big( \vec{k}, k_{\parallel}^{\prime} \big) \right] \nonumber\\
& \times & \int_{0}^{\infty} d t \; \xi \big( k_{\parallel}^{\prime}, t \big) e^{- \kappa_{\perp} k_{\perp}^2 t}.
\label{EUNLTafterdiffapprox}
\edm
Employing Eq. (\ref{formforWoldapp}) to replace $\xi ( k_{\parallel}^{\prime},t)$ allows us to solve the time-integral in Eq. (\ref{EUNLTafterdiffapprox}).
We find
\bdm
& & \int_{0}^{\infty} d t \; \xi \big( k_{\parallel}^{\prime}, t \big) e^{- \kappa_{\perp} k_{\perp}^2 t} \nonumber\\
& = & \frac{v^2}{3} \frac{1}{\omega_+ - \omega_-} \left[ \frac{\omega_{-}}{\omega_{-} - \kappa_{\perp} k_{\perp}^2} - \frac{\omega_{+}}{\omega_{+} - \kappa_{\perp} k_{\perp}^2} \right] \nonumber\\
& = & \frac{v^2}{3} \frac{\kappa_{\perp} k_{\perp}^2}{\big( \omega_{+} - \kappa_{\perp} k_{\perp}^2 \big)\big( \omega_{-} - \kappa_{\perp} k_{\perp}^2 \big)} \nonumber\\
& = & \frac{v^2}{3} \frac{\kappa_{\perp} k_{\perp}^2}{\omega_{+} \omega_{-} - \big( \omega_{+} + \omega_{-} \big) \kappa_{\perp} k_{\perp}^2 + \big( \kappa_{\perp} k_{\perp}^2 \big)^2}.
\label{perpwithomegas}
\edm
Using this in Eq. (\ref{EUNLTafterdiffapprox}) yields
\bdm
\kappa_{\perp}
& = & \frac{1}{2 \pi} \frac{1}{B_0^2} \frac{v^2}{3} \int_{-\infty}^{+\infty} d k_{\parallel}^{\prime} \nonumber\\
& \times & \int d^3 k \; P_{xx} \big( \vec{k} \big) \left[ R_{+} \big( \vec{k}, k_{\parallel}^{\prime} \big) + R_{-} \big( \vec{k}, k_{\parallel}^{\prime} \big) \right] \nonumber\\
& \times & \frac{\kappa_{\perp} k_{\perp}^2}{\omega_{+} \omega_{-} - \big( \omega_{+} + \omega_{-} \big) \kappa_{\perp} k_{\perp}^2 + \big( \kappa_{\perp} k_{\perp}^2 \big)^2}.
\edm
From Eq. (\ref{omegapm}) we can easily derive
\be
\omega_{+} + \omega_{-} = - v / \lambda_{\parallel}
\ee
as well as
\be
\omega_{+} \omega_{-} = \frac{1}{3} v^2 k_{\parallel}^{\prime \; 2}.
\ee
Using this in Eq. (\ref{perpwithomegas}) yields after some straightforward algebra
\bdm
\kappa_{\perp}
& = & \frac{1}{2 \pi} \frac{1}{B_0^2} \frac{v^2}{3} \int_{-\infty}^{+\infty} d k_{\parallel}^{\prime} \nonumber\\
& \times & \int d^3 k \; P_{xx} \big( \vec{k} \big) \left[ R_{+} \big( \vec{k}, k_{\parallel}^{\prime} \big) + R_{-} \big( \vec{k}, k_{\parallel}^{\prime} \big) \right] \nonumber\\
& \times & \frac{1}{v/\lambda_{\parallel} + v^2 k_{\parallel}^{\prime \; 2} / (3 \kappa_{\perp} k_{\perp}^2) + \kappa_{\perp} k_{\perp}^2}.
\label{diffEUNLT}
\edm
The latter equation corresponds to the diffusive version of Eq. (\ref{holyequation}). Therefore, we call Eq. (\ref{diffEUNLT}) the \textit{diffusive
FLPD theory}. It needs to be emphasized that Eq. (\ref{holyequation}) is more general and accurate since it is not based on this additional approximation.
However, Eq. (\ref{diffEUNLT}) is easier to apply and more convenient for finding pure analytical results.

\subsection{Small Kubo Number Turbulence}

We can simplify Eq. (\ref{diffEUNLT}) by considering small Kubo number turbulence. In this limit the resonance functions $R_{\pm}$ can be
replaced by using Eq. (\ref{smallKuboresonance}). After evaluating the $k_{\parallel}^{\prime}$-integral we then find
\bdm
\kappa_{\perp}
= \frac{v^2}{3 B_0^2} \int d^3 k \; \frac{P_{xx} \big( \vec{k} \big)}{v/\lambda_{\parallel} + v^2 k_{\parallel}^{2} / (3 \kappa_{\perp} k_{\perp}^2) + \kappa_{\perp} k_{\perp}^2}.\nonumber\\
\label{recdiffUNLT}
\edm
The latter formula agrees with diffusive UNLT theory (see Eq. (\ref{UNLT}) of the current paper). However, this agreement is only obtained
for turbulence with small Kubo numbers. Note, we could also derive Eq. (\ref{recdiffUNLT}) by directly combining the diffusion approximation
with Eq. (\ref{rederivetimeUNLT}). Diagram \ref{TheDiagram} shows these two paths from Eq. (\ref{holyequation}) to (\ref{recdiffUNLT}).

\subsection{Two-Dimensional Turbulence}

We now consider the case of two-dimensional turbulence but with diffusion approximation. In Eq. (\ref{diffEUNLT}) we can use the spectral tensor
corresponding to two-dimensional turbulence (\ref{Pnm2D}) to find
\bdm
\kappa_{\perp}
& = & \frac{v^2}{3 B_0^2} \int_{0}^{\infty} d k_{\perp} \; g^{2D} (k_{\perp}) \int_{-\infty}^{+\infty} d k_{\parallel} \; \frac{\kappafl k_{\perp}^{2}}{k_{\parallel}^{2} + \left( \kappafl k_{\perp}^{2} \right)^2} \nonumber\\
& \times & \frac{1}{v/\lambda_{\parallel} + v^2 k_{\parallel}^{2} / (3 \kappa_{\perp} k_{\perp}^2) + \kappa_{\perp} k_{\perp}^2}.
\label{kperpg2D1}
\edm
The $k_{\parallel}$-integral therein can be solved by (see, e.g., Gradshteyn \& Ryzhik (2000))
\be
\int_{-\infty}^{+\infty} d k_{\parallel} \; \frac{1}{k_{\parallel}^2 + b^2} \frac{1}{c k_{\parallel}^2 + d}
= \frac{\pi}{b} \frac{1}{b \sqrt{c d} + d}.
\ee
Using this in combination with $b = \kappafl k_{\perp}^{2}$, $c = v^2 / 3$, and
$d = v / \lambda_{\parallel} \kappa_{\perp} k_{\perp}^2 + \big( \kappa_{\perp} k_{\perp}^2 \big)^2$ allows us to write Eq. (\ref{kperpg2D1}) as
\bdm
\kappa_{\perp}
& = & \pi \frac{v^2}{3 B_0^2} \int_{0}^{\infty} d k_{\perp} \; g^{2D} (k_{\perp}) \nonumber\\
& \times & \frac{1}{\frac{v \kappafl}{\sqrt{3 \kappa_{\perp}}} k_{\perp} \sqrt{v / \lambda_{\parallel} + \kappa_{\perp} k_{\perp}^2} + v / \lambda_{\parallel} + \kappa_{\perp} k_{\perp}^2}.
\label{kperpg2D2}
\edm
After comparing this result with equations derived from previous theories (see, e.g., Eq. (\ref{NLGC}) of the current paper), we can see that the
two terms $v / \lambda_{\parallel}$ and $\kappa_{\perp} k_{\perp}^2$ are as before. If the term $v / \lambda_{\parallel}$ is dominant, this is in
particular the case in the formal limit $\lambda_{\parallel} \rightarrow 0$, the solution would be independent of the turbulence spectrum,
namely $\kappa_{\perp} / \kappa_{\parallel} = \delta B_x^2 / B_0^2$ corresponding to the fluid limit as given by Eq. (\ref{Fluid}). In this
case particles interact predominantly with ballistic magnetic field lines while their parallel motion is diffusive. If the term $\kappa_{\perp} k_{\perp}^2$
in Eq. (\ref{kperpg2D2}) is dominant, on the other hand, one finds the usual field line random walk limit where the perpendicular diffusion coefficient
is given by $\kappa_{\perp} \approx v \kappafl / 2$. As in some other results shown in this paper, the factor $4/3$ is also missing in front of the
$\kappa_{\perp} k_{\perp}^2$-term in Eq. (\ref{kperpg2D2}). As before, this is due to the approximations used to derive the parallel distribution function
in Fourier space. In Eq. (\ref{kperpg2D2}) we can clearly see that compared to previous descriptions of perpendicular transport, there is a new term
containing explicitly the field line diffusion coefficient $\kappafl$. In the next section we explore the physical meaning of this term.

\section{Detailed Analytical Discussion for 2D Turbulence}

In applications analytical forms of the perpendicular diffusion coefficient based on two-component turbulence with a dominant 2D contribution are frequently
used. This concerns studies of solar modulation (see, e.g., Moloto \& Engelbrecht (2020), Engelbrecht \& Wolmarans (2020), and Engelbrecht \& Moloto (2021))
as well diffusive shock acceleration (see, e.g., Zank et al. (2000), Li et al. (2003), Li et al. (2005), Zank et al. (2006), Li et al. (2012), and Ferrand et al. (2014)).
Therefore, we now focus on the new result obtained for two-dimensional turbulence after we have employed the diffusion approximation as given by Eq. (\ref{kperpg2D2}).

\subsection{Detailed Investigation of the New Term}

\begin{figure}
\centering
\includegraphics[width=0.48\textwidth]{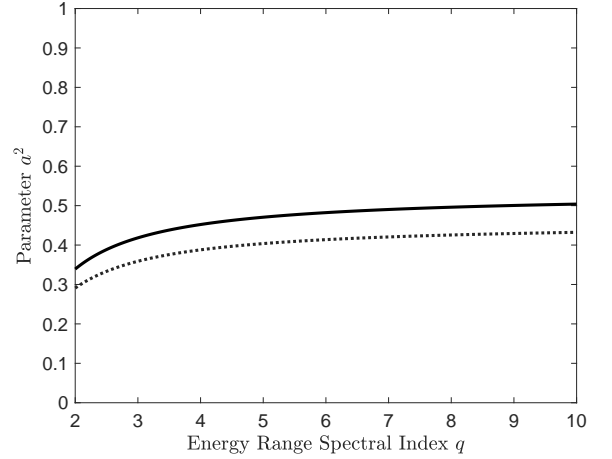}
\caption{The parameter $a^2$ as defined via Eq. (\ref{a2vsq}) versus the energy range spectral index $q$. Shown are the results for
$s=5/3$ (solid line) and $s=3/2$ (dotted line). Very clearly the parameter $a^2$ depends only weakly on the two spectral indices $s$ and $q$.}
\label{a2ersusq}
\end{figure}

In order to understand the physics of the new term in Eq. (\ref{kperpg2D2}) better, we consider the case of short parallel mean free paths but
assume that the term containing the field line diffusion coefficient is still dominant so that we approximate
\be
\frac{v \kappafl}{\sqrt{3 \kappa_{\perp}}} k_{\perp} \sqrt{\frac{v}{\lambda_{\parallel}} + \kappa_{\perp} k_{\perp}^2} + \frac{v}{\lambda_{\parallel}} + \kappa_{\perp} k_{\perp}^2 \approx \frac{v \kappafl}{\sqrt{3 \kappa_{\perp}}} k_{\perp} \sqrt{\frac{v}{\lambda_{\parallel}}}.
\ee
Using this approximation in Eq. (\ref{kperpg2D2}) yields
\bdm
\kappa_{\perp}
& = & \pi \frac{v^2}{3 B_0^2} \int_{0}^{\infty} d k_{\perp} \; \frac{g^{2D} (k_{\perp})}{\frac{v \kappafl}{\sqrt{3 \kappa_{\perp}}} k_{\perp} \sqrt{v / \lambda_{\parallel}}} \nonumber\\
& = & \pi \frac{v^2}{3 B_0^2} \frac{\sqrt{3 \kappa_{\perp}}}{v \kappafl} \sqrt{\frac{\lambda_{\parallel}}{v}}
\int_{0}^{\infty} d k_{\perp} \; g^{2D} (k_{\perp}) k_{\perp}^{-1}.
\label{kperpg2D3}
\edm
The remaining integral could easily be evaluated for a certain spectrum. In the following we want to keep our investigations as general as possible.
Therefore, we replace the $k_{\perp}$-integral by the so-called perpendicular integral scale of the turbulence $L_{\perp}$ which is defined
via (see, e.g., Shalchi (2020a))
\be
\delta B_x^2 L_{\perp} = 2 \pi \int_{0}^{\infty} d k_{\perp} \; g^{2D} (k_{\perp}) k_{\perp}^{-1}.
\ee
Therewith, Eq. (\ref{kperpg2D3}) can be written as
\be
\kappa_{\perp} = \frac{v}{6 B_0^2} \frac{\sqrt{3 \kappa_{\perp}}}{\kappafl} \sqrt{\frac{\lambda_{\parallel}}{v}} L_{\perp} \delta B_x^2.
\label{kperpg2D4}
\ee
To simplify this further we replace the parallel mean free path by the parallel spatial diffusion coefficient via $\lambda_{\parallel} = 3 \kappa_{\parallel} / v$
to find
\be
\sqrt{\frac{\kappa_{\perp}}{\kappa_{\parallel}}} = \frac{1}{2} \frac{L_{\perp}}{\kappafl} \frac{\delta B_x^2}{B_0^2}.
\label{sqrtratio}
\ee
The latter result can be rewritten in different ways. First we can replace the field line diffusion coefficient by the two-dimensional
result given by Eq. (\ref{kappaflLU}). Using this and taking the square of Eq. (\ref{sqrtratio}) gives us
\be
\frac{\kappa_{\perp}}{\kappa_{\parallel}} = \frac{L_{\perp}^2}{4 L_U^2} \frac{\delta B_x^2}{B_0^2}
\label{kperpg2D5}
\ee
where we have used again the ultra-scale $L_U$. In order to evaluate this in more detail, we now need to specify the turbulence spectrum.
In what follows we use a general spectrum with energy and inertial range. Those two parts of the spectrum are separated via the
characteristic scale $\ell_{\perp}$. Such a spectrum was developed in Shalchi \& Weinhorst (2009) which has the following form
\be
g^{2D}(k_{\perp}) = \frac{4 D(s, q)}{\pi} \delta B_{x}^2 \ell_{\perp}
\frac{(k_{\perp} \ell_{\perp})^{q}}{\left[ 1 + (k_{\perp} \ell_{\perp})^2 \right]^{(s+q)/2}}.
\label{2dspec}
\ee
In the inertial range the spectrum scales like $k_{\perp}^{-s}$ whereas in the energy range it scales like $k_{\perp}^{q}$. In Eq. (\ref{2dspec})
we have also used the normalization function
\be
D(s, q) = \frac{\Gamma \left( \frac{s+q}{2} \right)}{2 \Gamma \left( \frac{s-1}{2} \right) \Gamma \left( \frac{q+1}{2} \right)}.
\label{normalD}
\ee
For most calculations performed in the current paper involving two-dimensional modes, we have set $s=5/3$ and $q=3$. As, for instance,
demonstrated in Shalchi (2020a), the perpendicular integral scale is for this specific spectrum
\be
L_{\perp}
= \frac{2 \Gamma \left( \frac{q}{2} \right) \Gamma \left( \frac{s}{2} \right)}{\Gamma \left( \frac{q+1}{2} \right) \Gamma \left( \frac{s-1}{2} \right)} \ell_{\perp}
\label{LperpWeinhorst}
\ee
and the ultra-scale is given by
\be
L_U = \sqrt{\frac{s-1}{q-1}} \ell_{\perp}.
\label{LUforWeinhorst}
\ee
Analytical results derived in the past based on NLGC and UNLT theories yield for short parallel mean free paths (see, e.g., Shalchi et al. (2004),
Zank et al. (2004), and Shalchi et al. (2010)) 
\be
\frac{\kappa_{\perp}}{\kappa_{\parallel}} = a^2 \frac{\delta B_x^2}{B_0^2}.
\ee
Comparing this with Eq. (\ref{kperpg2D5}) allows us to write
\be
a^2 = \frac{L_{\perp}^2}{4 L_U^2}.
\ee
According to the formulas listed above this becomes
\be
a^2 = \frac{q-1}{s-1} \left[ \frac{\Gamma \left( \frac{q}{2} \right) \Gamma \left( \frac{s}{2} \right)}{\Gamma \left( \frac{q+1}{2} \right) \Gamma \left( \frac{s-1}{2} \right)} \right]^2
\label{a2vsq}
\ee
meaning that for the used spectrum the parameter $a^2$ depends only on the two spectral indices $s$ and $q$. It needs to be emphasized
that only for $q>1$ are the aforementioned turbulence scales finite. This conclusion was originally obtained by Matthaeus et al. (2007)
for a slightly different spectrum. Eq. (\ref{a2vsq}) is visualized in Fig. \ref{a2ersusq} indicating that $a^2$ depends only weakly on
$s$ as well as $q$ and that $0.3 \leq a^2 \leq 0.5$.

Alternatively, Eq. (\ref{kperpg2D5}) can be written as
\be
\frac{\kappa_{\perp}}{\kappa_{\parallel}} = \frac{L_{\perp}^2}{4 L_U^4} \kappafl^2.
\label{obtainCLRR1}
\ee
In order to understand the latter formula, we need to replace the scales $L_{\perp}$ and $L_U$. As before, we employ Eqs. (\ref{LperpWeinhorst})
and (\ref{LperpWeinhorst}). Therewith, we can write Eq. (\ref{obtainCLRR1}) as
\be
\frac{\kappa_{\perp}}{\kappa_{\parallel}} = b^2 \frac{\kappafl^2}{\ell_{\perp}^2}
\label{obtainCLRR2}
\ee
where we have used
\be
b^2 = \frac{(q-1)^2}{(s-1)^2}
\left[ \frac{\Gamma \left( \frac{q}{2} \right) \Gamma \left( \frac{s}{2} \right)}{\Gamma \left( \frac{q+1}{2} \right) \Gamma \left( \frac{s-1}{2} \right)} \right]^2.
\ee
Eq. (\ref{obtainCLRR2}) is similar compared to the heuristic result discussed in the introduction (see Eq. (\ref{CLRR}) of the current paper).
Therefore, we can easily understand the physics of the new term found here. Particles follow diffusive field lines while their parallel motion
is diffusive as well. In this case we find compound sub-diffusion. As soon as the transverse complexity of the turbulence becomes important,
particles leave the original field line they followed and diffusion is restored. The obtained diffusion coefficient is comparable to the
CLRR limit.

\subsection{The Limit of Short Parallel Mean Free Paths}

So far we have explored the new term and neglected the previously obtained terms in Eq. (\ref{kperpg2D2}). However, there is a problem with those
considerations and that is that there is a \textit{competition} between two terms in the denominator of Eq. (\ref{kperpg2D2}). In the formal limit
$\lambda_{\parallel} \rightarrow 0$ the latter formula can be simplified to
\be
\frac{\kappa_{\perp}}{\kappa_{\parallel}} = \frac{\pi}{B_0^2} \int_{0}^{\infty} d k_{\perp} \;
\frac{g^{2D} (k_{\perp})}{1 + \kappafl k_{\perp} \sqrt{\kappa_{\parallel} / \kappa_{\perp}}}.
\label{shortparallel}
\ee
To be more specific we use spectrum (\ref{2dspec}) and employ the transformation $x = k_{\perp} \ell_{\perp}$ so that Eq. (\ref{shortparallel}) becomes
\be
\frac{\kappa_{\perp}}{\kappa_{\parallel}} = 4 D \big( s, q \big) \frac{\delta B_x^2}{B_0^2} \int_{0}^{\infty} d x \;
\frac{h \big( x \big)}{1 + \alpha x}
\label{CLRRfluidboth}
\ee
where we have used the dimensionless spectrum
\be
h \big( x \big) = \frac{x^q}{\left( 1 + x^2 \right)^{(s+q)/2}}
\label{definehx}
\ee
and the parameter
\be
\alpha := \frac{\kappafl}{\ell_{\perp}} \sqrt{\frac{\lambda_{\parallel}}{\lambda_{\perp}}}
\equiv \frac{\kappafl}{\ell_{\perp}} \sqrt{\frac{\kappa_{\parallel}}{\kappa_{\perp}}}.
\label{definealpha}
\ee
It follows from Eq. (\ref{CLRRfluidboth}) that we find the fluid limit (see Eq. (\ref{Fluid})) if $\alpha \ll 1$ and CLRR diffusion for $\alpha \gg 1$.
Therefore, the critical value, where the turnover from the fluid limit to CLRR diffusion occurs, is $\alpha \approx 1$. According to Eq. (\ref{definealpha}),
this condition can be written as
\be
\frac{\kappa_{\perp}}{\kappa_{\parallel}} = \left( \frac{\kappafl}{\ell_{\perp}} \right)^2 \quad\quad \textnormal{for} \quad\quad \alpha = 1
\ee
which actually corresponds to CLRR diffusion (see Eq. (\ref{CLRR}) of the current paper). However, both sides of the latter equation are always
directly proportional to $\delta B_{x}^2 / B_0^2$ and, therefore, changing the magnetic field ratio does not restore the fluid limit. We conclude
that real perpendicular diffusion, for the case of very short parallel mean free paths and large Kubo numbers, is always a combination of the fluid
limit and CLRR diffusion. The physics behind this is not too difficult to understand. The fluid limit is valid in cases where parallel transport
becomes diffusive almost instantaneously and particles interact with ballistic field lines before transverse complexity becomes important.
CLRR diffusion, on the other hand, is obtained if parallel transport becomes diffusive almost instantaneously, then the field lines become
diffusive and thereafter transverse complexity becomes significant. However, for large Kubo number turbulence such as quasi-2D turbulence,
field lines becomes diffusive exactly for the scales where transverse complexity becomes relevant, namely
$\langle ( \Delta x )^2 \rangle \approx 2 \ell_{\perp}^2$. Eq. (\ref{shortparallel}) takes into account all those effects and weights
them appropriately. Therefore, for three-dimensional turbulence with large Kubo numbers, including two-dimensional turbulence, and short
parallel mean free paths, we are in a hybrid regime. Fig. \ref{EUNLTRegimes} shows the different regimes of perpendicular diffusion.

\subsection{The Limit of Long Parallel Mean Free Paths}

\begin{figure}[ht]
\centering
\includegraphics[width=0.48\textwidth]{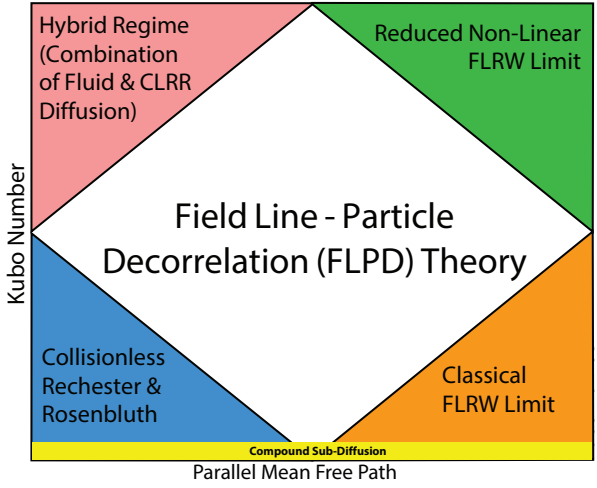}
\caption{Perpendicular diffusion of energetic particles in three-dimensional turbulence depends on two parameters, namely the Kubo number
given by Eq. (\ref{defineKuboNumber}) and the parallel mean free path. The shown diagram is an updated version compared to the diagram
depicted in Shalchi (2020a). For short parallel mean free paths and large Kubo numbers we find that particle transport is a combination
of CLRR (collisionless Rechester \& Rosenbluth) diffusion given by Eq. (\ref{CLRR}) and the fluid limit given by Eq. (\ref{Fluid}). This
is a highly non-linear regime where magnetic field lines show non-linear behavior and perpendicular transport is strongly influenced by
parallel transport. For long parallel mean free paths and large Kubo numbers we find a reduced FLRW (field line random walk) limit
where the field lines are still in the highly non-linear regime. The other two limits are as described by diffusive UNLT theory as 
discussed in Shalchi (2015) and Shalchi (2020a).}
\label{EUNLTRegimes}
\end{figure}

So far we focused on short parallel mean free paths. In the formal limit $\lambda_{\parallel} \rightarrow \infty$, on the other hand,
Eq. (\ref{kperpg2D2}) can be simplified to
\be
\kappa_{\perp} = \frac{v^2}{3} \frac{1}{\frac{v \kappafl}{\sqrt{3}} + \kappa_{\perp}} \frac{\pi}{B_0^2}
\int_{0}^{\infty} d k_{\perp} \; g^{2D} (k_{\perp}) k_{\perp}^{-2}.
\label{longlimit1}
\ee
The field line diffusion coefficient for quasi-2D turbulence is given by (see Matthaeus et al. (1995))
\be
\kappafl^2 = \frac{\pi}{B_0^2} \int_{0}^{\infty} d k_{\perp} \; g^{2D} (k_{\perp}) k_{\perp}^{-2}
\ee
and, therefore, Eq. (\ref{longlimit1}) can simply be written as
\be
\kappa_{\perp} = \frac{v^2}{3} \frac{\kappafl^2}{v \kappafl / \sqrt{3} + \kappa_{\perp}}.
\label{longlimit2}
\ee
The latter formula corresponds to the quadratic equation
\be
\kappa_{\perp}^2 + \frac{v \kappafl}{\sqrt{3}} \kappa_{\perp} - \frac{v^2}{3} \kappafl^2 = 0.
\ee
The only physical solution of the latter equation is
\be
\kappa_{\perp} = \frac{\sqrt{5} - 1}{\sqrt{3}} \frac{v}{2} \kappafl \approx 0.7136 \; \frac{v}{2} \kappafl.
\label{reducedFLRW}
\ee
Compared to the real field line random walk limit, which is given by $\kappa_{\perp} = v \kappafl / 2$, our new result is reduced
by a factor $1.5$. This can also be understood. The real field line random walk limit is obtained only if parallel diffusion is suppressed
forever and if particles follows field lines forever. Both assumption are unrealistic. In reality particle transport along the mean field
always becomes diffusive if time passes and particles always need to leave the original field line they were tied to. However, in the limit
of very long parallel mean free paths, perpendicular transport is still predominantly controlled by the field line random walk because
pitch-angle scattering is in this case a weak process. Fig. \ref{EUNLTRegimes} depicts the different transport regimes contained in FLPD theory.

\section{Comparison with Test-Particle Simulations}

A significant amount of test-particle simulations was performed in the past for a slab+2D turbulence model also known as two-component turbulence
(see, e.g., Qin et al. (2002a,b), Qin \& Shalchi (2012), Hussein et al. (2015), and Arendt \& Shalchi (2020)). The two-component model is strongly
supported by solar wind observations where one can find a cross distribution of magnetic fluctuations (see Matthaeus et al. (1990)) indicating
that real turbulence can be approximated by a superposition of slab and two-dimensional modes. Furthermore, as demonstrated by
Bieber et al. (1996), solar wind turbulence possesses a dominant (roughly $80 \%$ by magnetic energy) two-dimensional component and a
smaller (roughly $20 \%$) slab contribution. Theoretically, the strong two-dimensional component was explained based on the theory
of nearly incompressible magnetohydrodynamics (see Zank \& Matthaeus (1993)). In the recent years those conclusions have been confirmed
(see, e.g., Zank et al. (2017) and Zank et al. (2020)). In the following we employ the two-component turbulence model and compute perpendicular
diffusion coefficients based on the theory developed in the current paper. We then compare our analytical findings with previously performed
test-particle simulations.

The explicit slab contribution to the perpendicular diffusion coefficient is zero. For two-dimensional turbulence we derived Eq. (\ref{kperpg2D2}).
In the following we rewrite this to make it more convenient for numerical investigations. Replacing the perpendicular diffusion coefficient by the
perpendicular mean free path via $\kappa_{\perp} = v \lambda_{\perp} / 3$ allows us to write Eq. (\ref{kperpg2D2}) as
\bdm
\frac{\lambda_{\perp}}{\lambda_{\parallel}}
& = & \frac{\pi}{B_0^2} \int_{0}^{\infty} d k_{\perp} \; g^{2D} (k_{\perp}) \nonumber\\
& \times & \frac{1}{\kappafl k_{\perp} \sqrt{\frac{\lambda_{\parallel}}{\lambda_{\perp}}} \sqrt{1 + \frac{1}{3} \lambda_{\parallel} \lambda_{\perp} k_{\perp}^2}
+ 1 + \frac{1}{3} \lambda_{\parallel} \lambda_{\perp} k_{\perp}^2}.\nonumber\\
\label{kappafor2d}
\edm
For the spectral function associated with the two-dimensional modes we employ again Eq. (\ref{2dspec}). After using this in Eq. (\ref{kappafor2d})
and performing the integral-transformation $x = k_{\perp} \ell_{\perp}$ we find
\bdm
\frac{\lambda_{\perp}}{\lambda_{\parallel}}
& = & 4 D(s, q) \frac{\delta B_x^2}{B_0^2} \nonumber\\
& \times & \int_{0}^{\infty} d x \;
\frac{h(x)}{\frac{\kappafl}{\ell_{\perp}} \sqrt{\frac{\lambda_{\parallel}}{\lambda_{\perp}}} x \sqrt{1 + \frac{\lambda_{\parallel} \lambda_{\perp}}{3 \ell_{\perp}^2} x^2} + 1 + \frac{\lambda_{\parallel} \lambda_{\perp}}{3 \ell_{\perp}^2} x^2}\nonumber\\
\label{dimlessEUNLT}
\edm
where we have used the dimensionless spectrum given by Eq. (\ref{definehx}).

A peculiarity of FLPD theory is that it explicitly contains the field line diffusion coefficient $\kappafl$. The total field line diffusion coefficient
in two-component turbulence is (see Matthaeus et al. (1995))
\be
\kappafl = \frac{1}{2} \left[ \kappa_{slab} + \sqrt{\kappa_{slab}^2 + 4 \kappa_{2D}^2} \right]
\ee
which depends on the slab as well as the two-dimensional diffusion coefficients. However, a problem of FLPD theory is that it is not clear whether
the parameter $\kappafl$ in Eq. (\ref{dimlessEUNLT}) is the total field line diffusion coefficient or the field line diffusion coefficient
associated with the two-dimensional modes $\kappa_{2D}$. In the following we assume that it is the total diffusion coefficient but further
investigations need to explore whether this is true or not.

In order to determine the individual coefficients $\kappa_{slab}$ and $\kappa_{2D}$, we need to specify the two spectra. For the slab modes we
use the Bieber et al. (1994) model which is given by
\be
g^{slab} (k_{\parallel}) = \frac{1}{2\pi} C(s)\delta B_{slab}^2 \ell_\parallel \left[ 1 + \left( k_\parallel \ell_\parallel \right)^2 \right]^{-s/2}
\label{SlabSpectrum}
\ee
where we have used the normalization function
\be
C(s) = \frac{\Gamma \left( \frac{s}{2} \right)}{2 \sqrt{\pi} \Gamma \left( \frac{s-1}{2} \right)}
\label{normalC}
\ee
where $s$ is the inertial range spectral index as before and $\ell_\parallel$ is the parallel (or slab) bendover scale. For the
two-dimensional modes we employ the form given by Eq. (\ref{2dspec}). For those spectra, the slab field line diffusion coefficient
is given by
\be
\kappa_{slab} = L_{\parallel} \frac{\delta B_x^2}{B_0^2} = \pi C(s) \ell_{\parallel} \frac{\delta B_{slab}^2}{B_0^2}
\ee
and the two-dimensional field line diffusion coefficient is given by (see, e.g., Shalchi (2020a) for a systematic derivation of such results)
\be
\kappa_{2D} = L_{U} \frac{\delta B_x}{B_0} = \sqrt{\frac{s-1}{2(q-1)}} \frac{\delta B_{2D}}{B_0}.
\ee
Eq. (\ref{dimlessEUNLT}) can easily be solved numerically. In fact it is not more difficult to solve this equation than the previous equations
derived in Matthaeus et al. (2003) or Shalchi (2010) (see, e.g., Eqs. (\ref{UNLT}) and (\ref{NLGC}) of the current paper).

\begin{figure*}[!tbp]
\centering
\begin{minipage}[b]{0.48\textwidth}
\includegraphics[width=\textwidth]{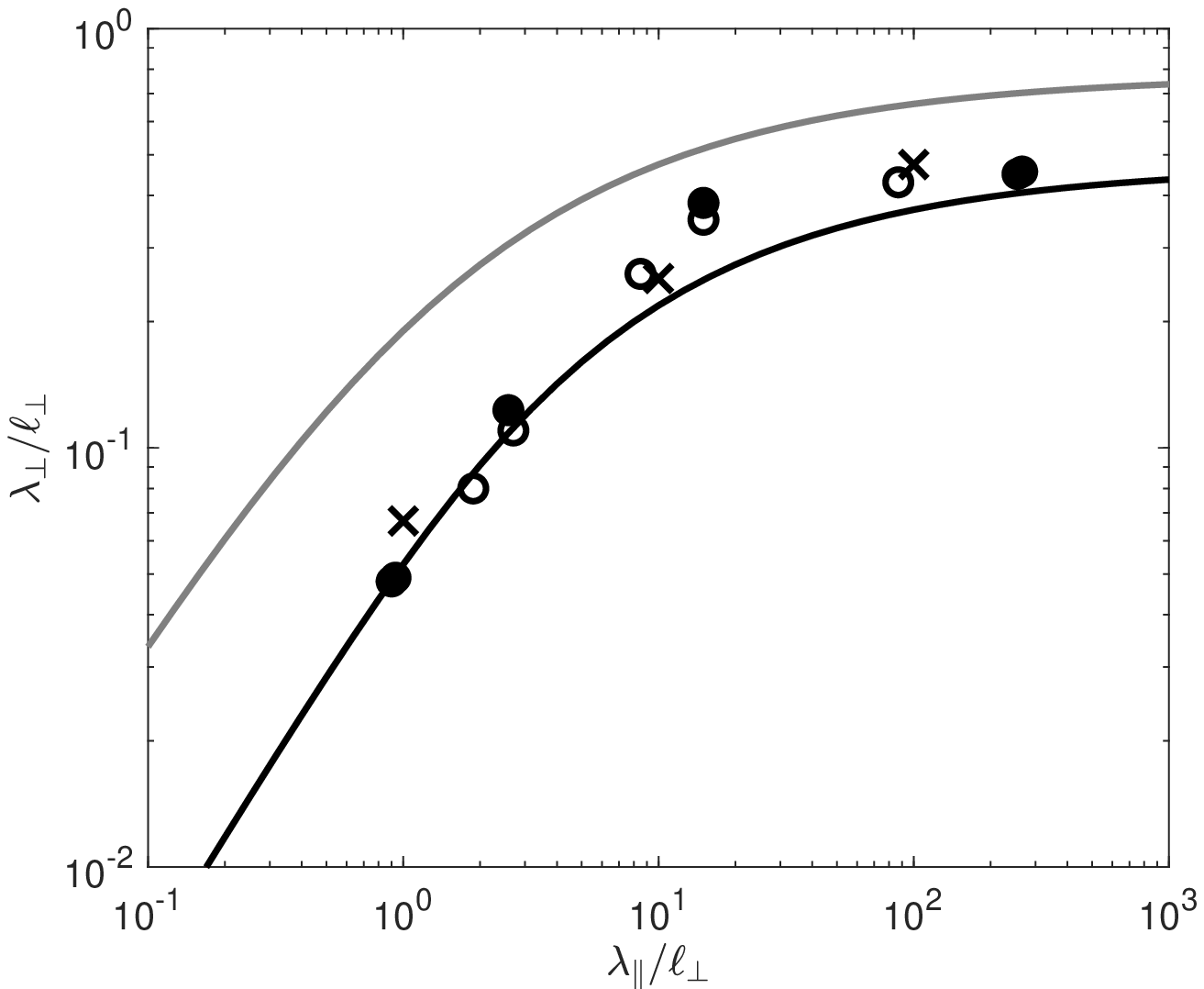}
\end{minipage}
\hfill
\begin{minipage}[b]{0.48\textwidth}
\includegraphics[width=\textwidth]{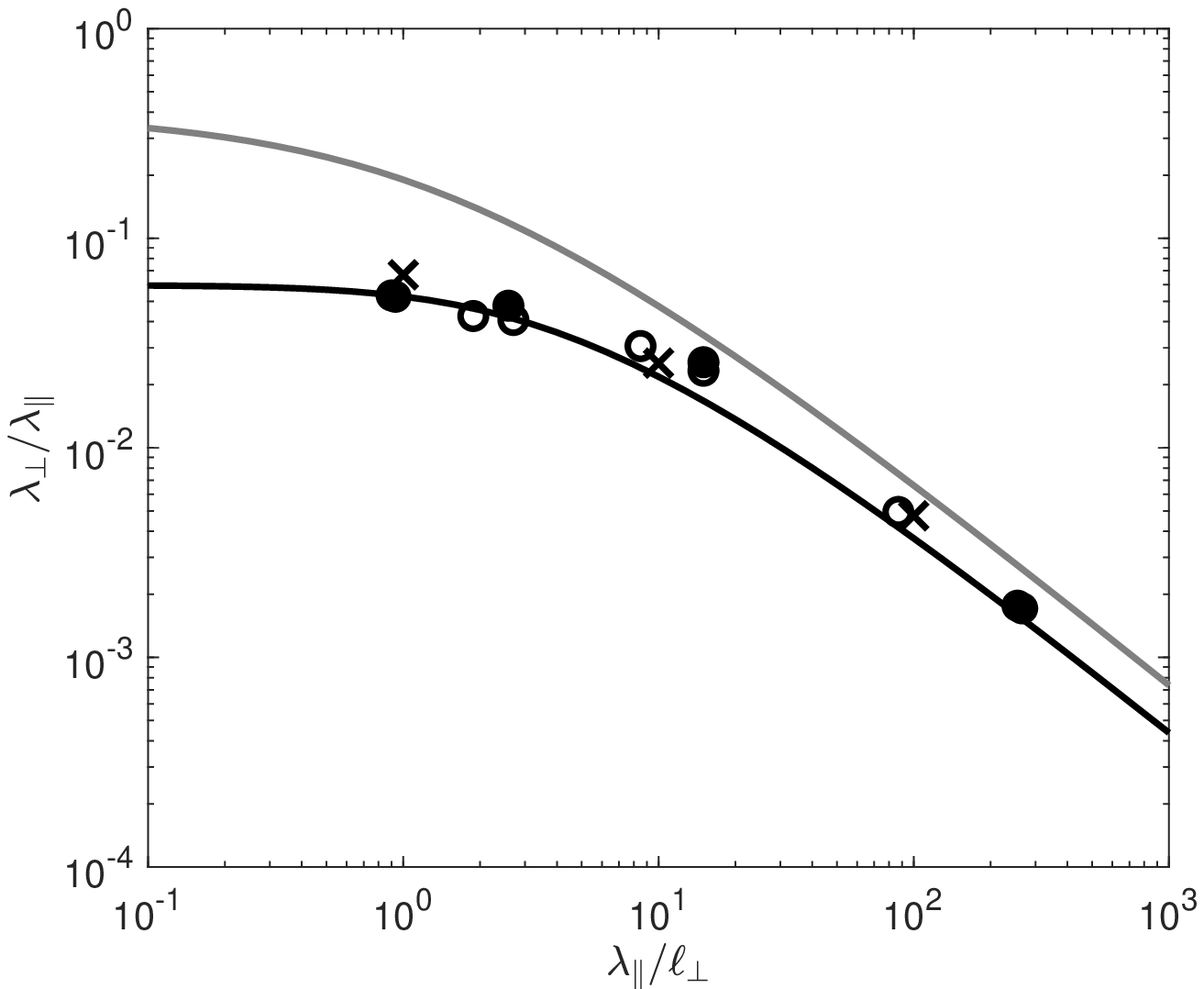}
\end{minipage}
\caption{The left panel shows the perpendicular mean free path versus the parallel mean free path and the right panel shows the ratio
$\lambda_{\perp} / \lambda_{\parallel} \equiv \kappa_{\perp} / \kappa_{\parallel}$ versus the parallel mean free path. The black solid lines
correspond to the diffusive FLPD results as given by Eq. (\ref{dimlessEUNLT}) and the circles represent the different test-particle simulations
published in Hussein et al. (2015). The crosses represent time-dependent FLPD theory as given by Eq. (\ref{holyequation}) or, in code units, by
Eq. (\ref{slab2Dfornumerics}). For comparison we have also shown the NLGC/UNLT results without correction factor, meaning that we have
set $a^2 = 1$ (gray solid line). To obtain the visualized results the following parameter values have been used: $\delta B^2 / B_0^2 = 1$,
$\ell_{\perp} = \ell_{\parallel}$, $q=2$, and a slab fraction of $20 \%$.}
\label{SimComp1}
\end{figure*}

\begin{figure*}[!tbp]
\centering
\begin{minipage}[b]{0.48\textwidth}
\includegraphics[width=\textwidth]{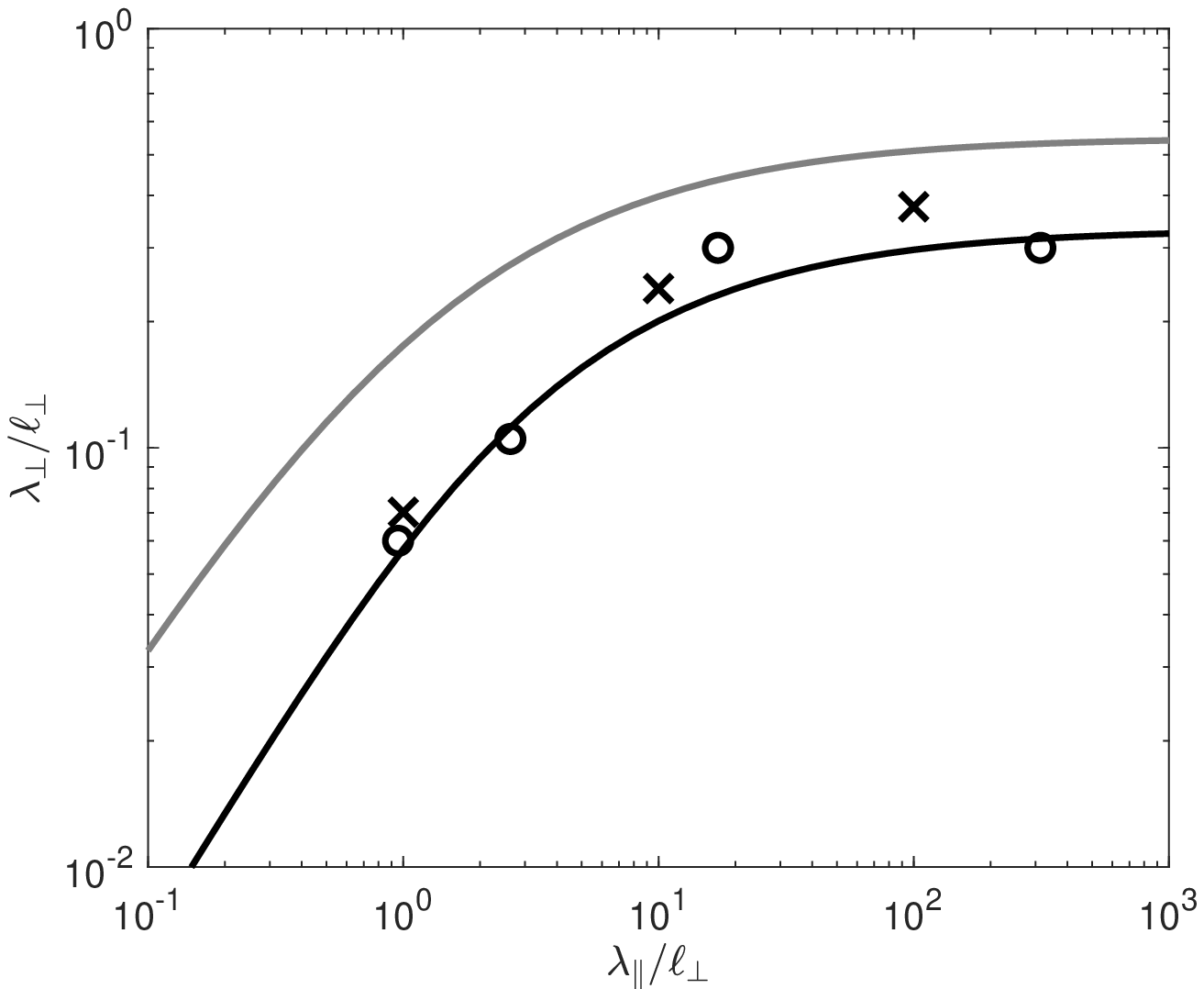}
\end{minipage}
\hfill
\begin{minipage}[b]{0.48\textwidth}
\includegraphics[width=\textwidth]{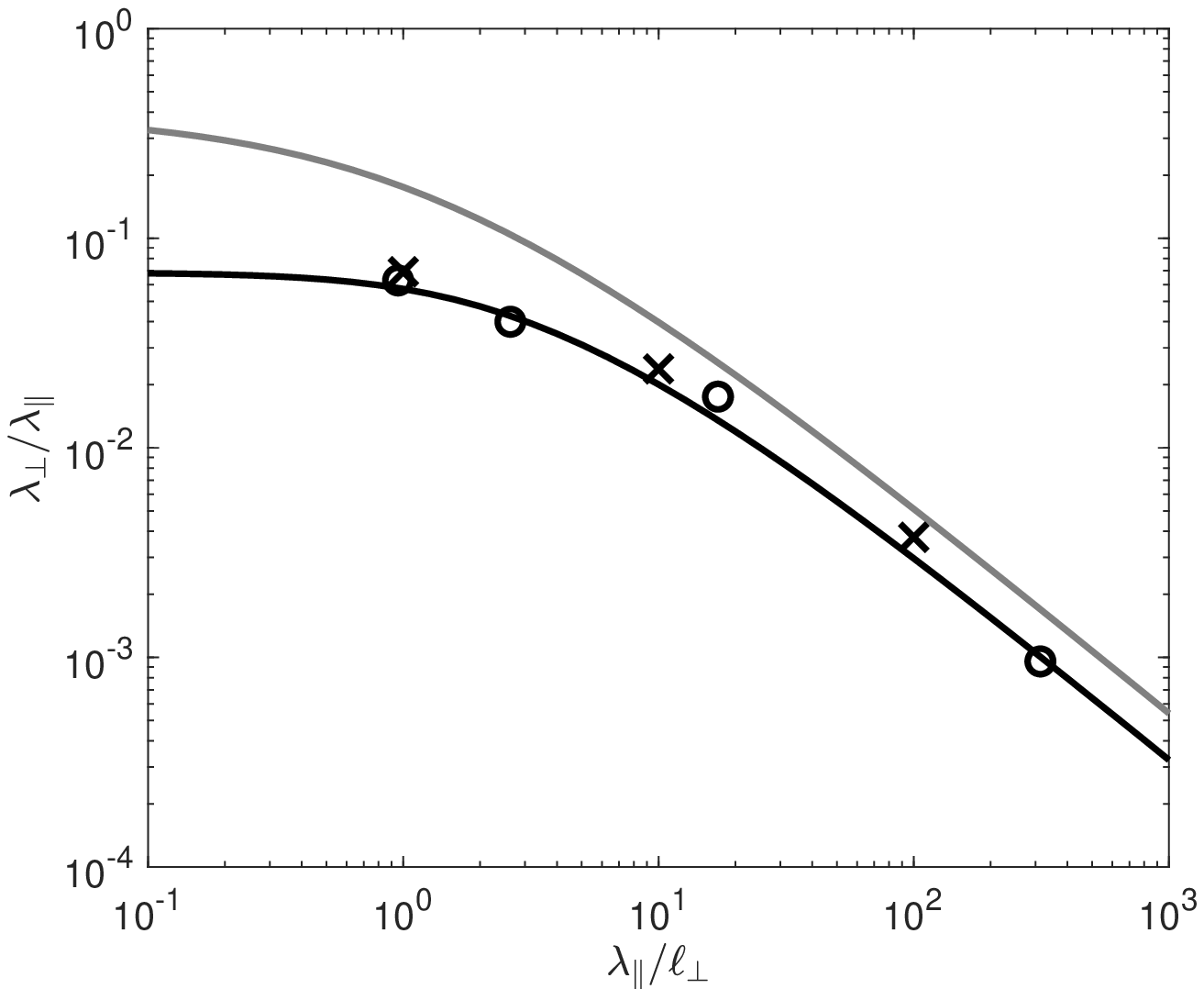}
\end{minipage}
\caption{The left panel shows the perpendicular mean free path versus the parallel mean free path and the right panel shows the ratio
$\lambda_{\perp} / \lambda_{\parallel} \equiv \kappa_{\perp} / \kappa_{\parallel}$ versus the parallel mean free path. The black solid lines
corresponds to the diffusive FLPD results as given by Eq. (\ref{dimlessEUNLT}) and the circles represent the test-particle simulations published
in Arendt \& Shalchi (2020). The crosses represent time-dependent FLPD theory as given by Eq. (\ref{holyequation}) or, in code units, by
Eq. (\ref{slab2Dfornumerics}). For comparison we have also shown the NLGC/UNLT results without correction factor, meaning that we have set $a^2 = 1$
(gray solid line). To obtain the visualized results the following parameter values have been used: $\delta B^2 / B_0^2 = 1$,
$\ell_{\perp} = \ell_{\parallel}$, $q=3$, and a slab fraction of $20 \%$.}
\label{SimComp2}
\end{figure*}

\begin{figure*}[!tbp]
\centering
\begin{minipage}[b]{0.48\textwidth}
\includegraphics[width=\textwidth]{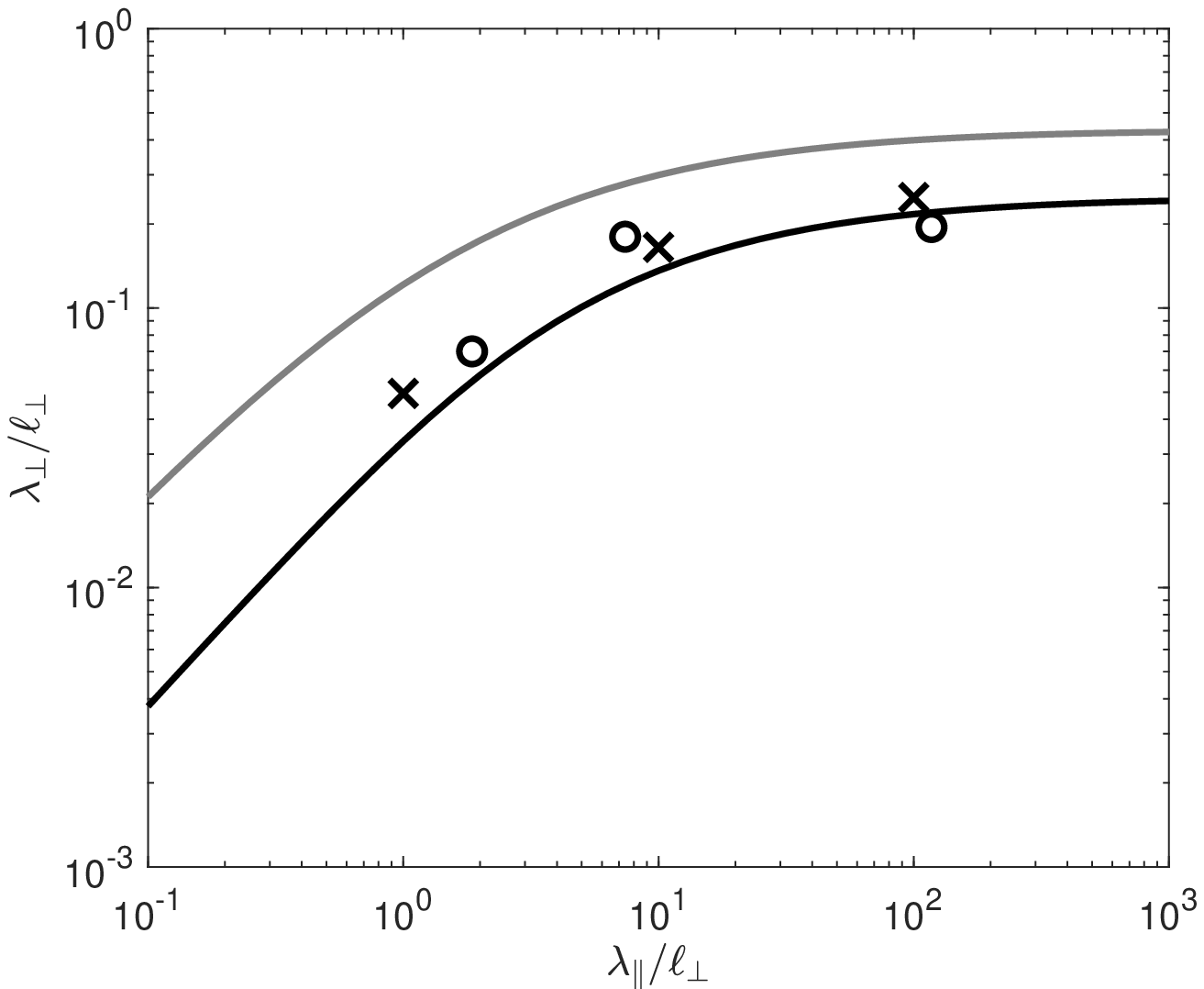}
\end{minipage}
\hfill
\begin{minipage}[b]{0.48\textwidth}
\includegraphics[width=\textwidth]{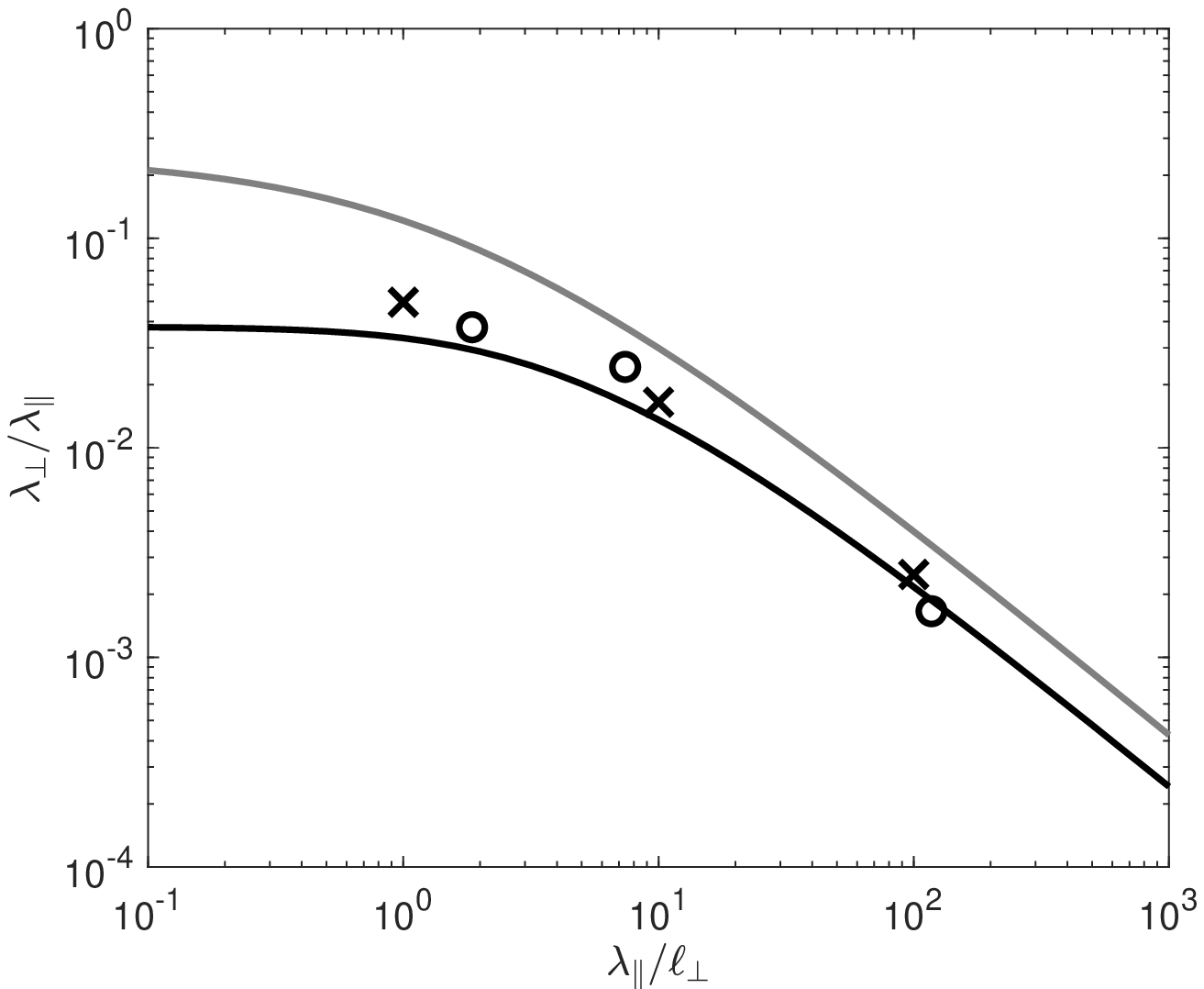}
\end{minipage}
\caption{Caption is as in Fig. \ref{SimComp2} but the following parameter values have been used: $\delta B^2 / B_0^2 = 1$,
$\ell_{\perp} = \ell_{\parallel}$, $q=3$, and a slab fraction of $50 \%$.}
\label{SimComp3}
\end{figure*}

\begin{figure*}[!tbp]
\centering
\begin{minipage}[b]{0.48\textwidth}
\includegraphics[width=\textwidth]{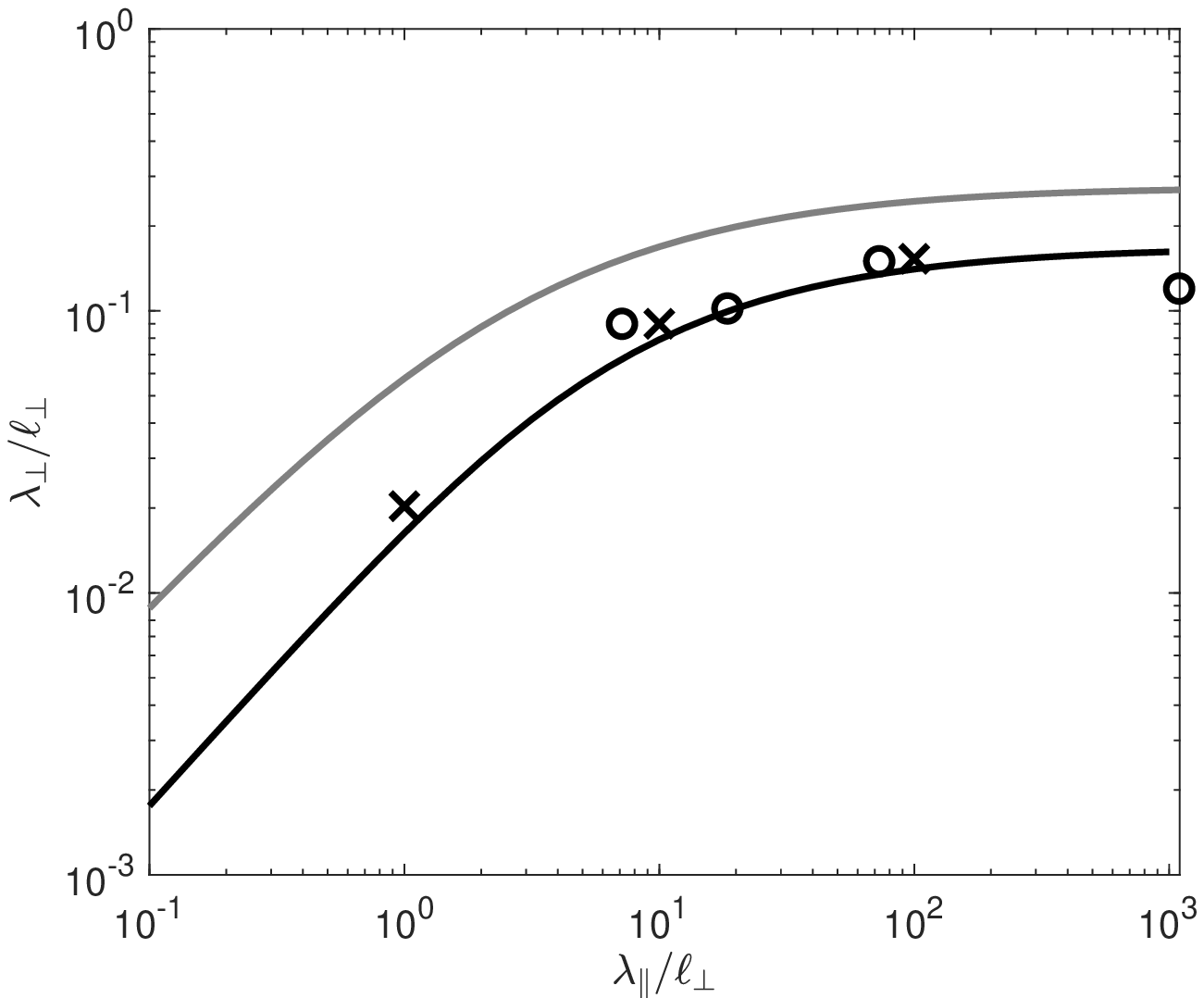}
\end{minipage}
\hfill
\begin{minipage}[b]{0.48\textwidth}
\includegraphics[width=\textwidth]{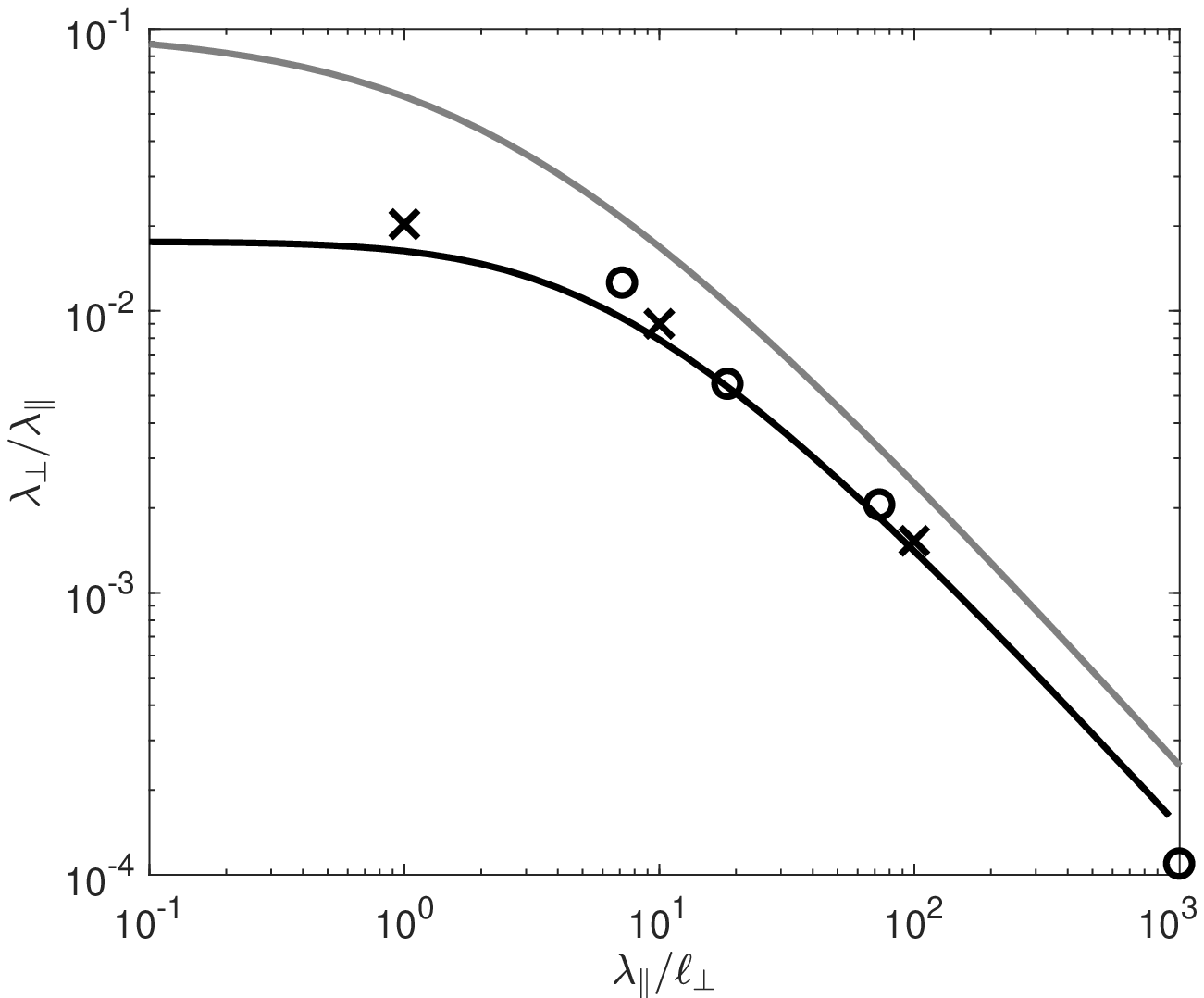}
\end{minipage}
\caption{Caption is as in Fig. \ref{SimComp2} but the following parameter values have been used: $\delta B^2 / B_0^2 = 0.25$,
$\ell_{\perp} = \ell_{\parallel}$, $q=3$, and a slab fraction of $20 \%$.}
\label{SimComp4}
\end{figure*}

\begin{figure*}[!tbp]
\centering
\begin{minipage}[b]{0.48\textwidth}
\includegraphics[width=\textwidth]{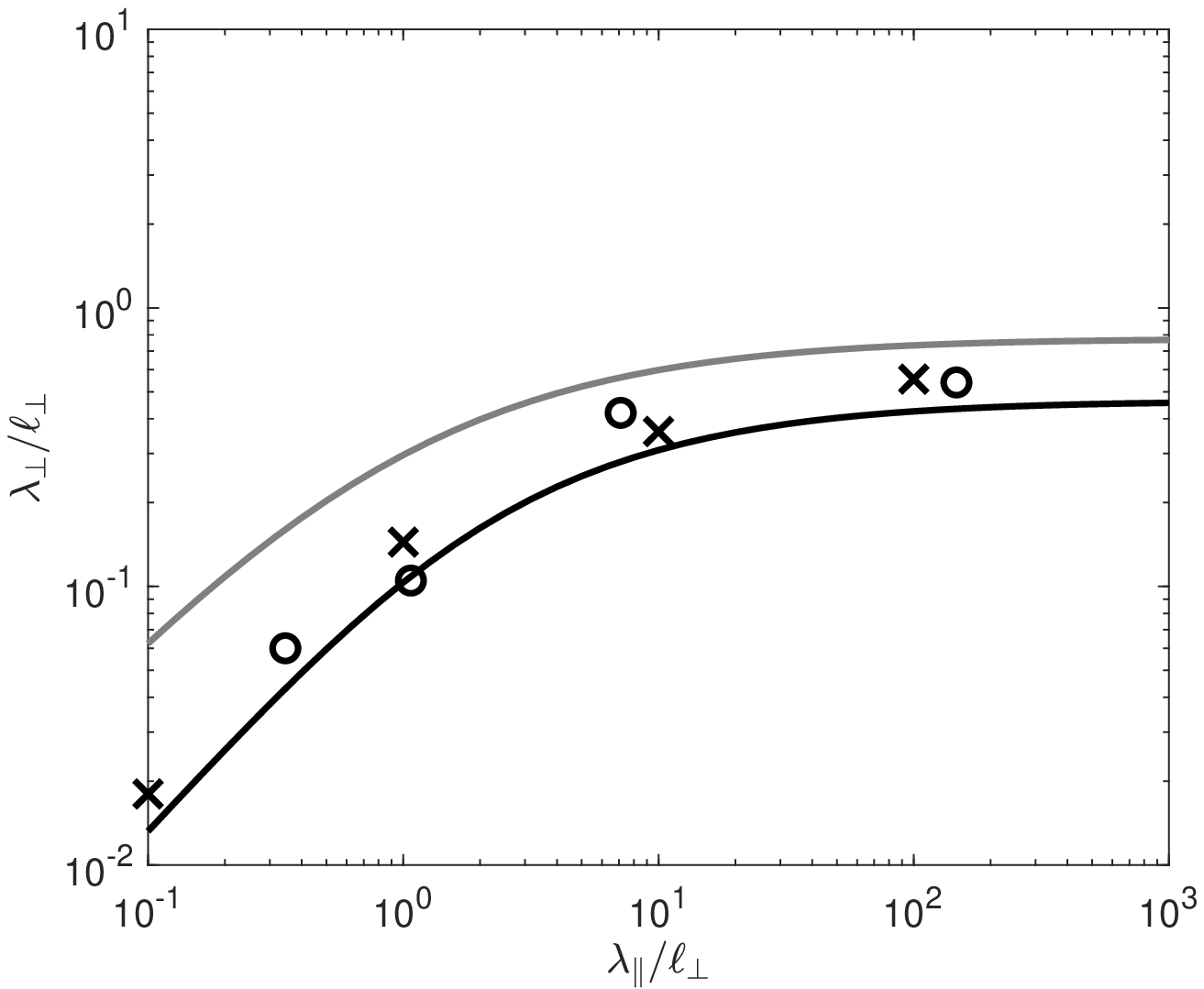}
\end{minipage}
\hfill
\begin{minipage}[b]{0.48\textwidth}
\includegraphics[width=\textwidth]{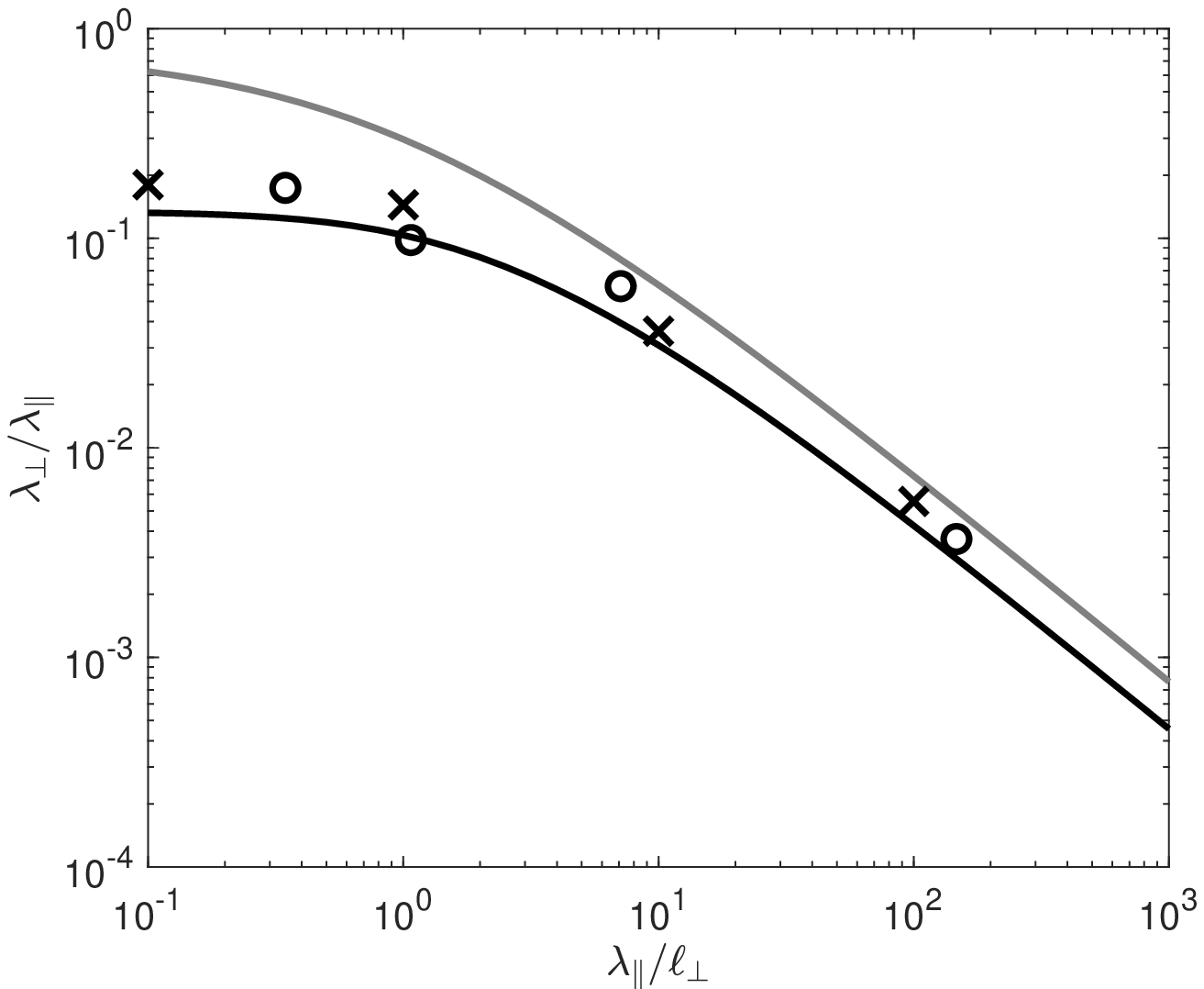}
\end{minipage}
\caption{Caption is as in Fig. \ref{SimComp2} but the following parameter values have been used: $\delta B^2 / B_0^2 = 2$,
$\ell_{\perp} = \ell_{\parallel}$, $q=3$, and a slab fraction of $20 \%$.}
\label{SimComp5}
\end{figure*}

\begin{figure*}[!tbp]
\centering
\begin{minipage}[b]{0.48\textwidth}
\includegraphics[width=\textwidth]{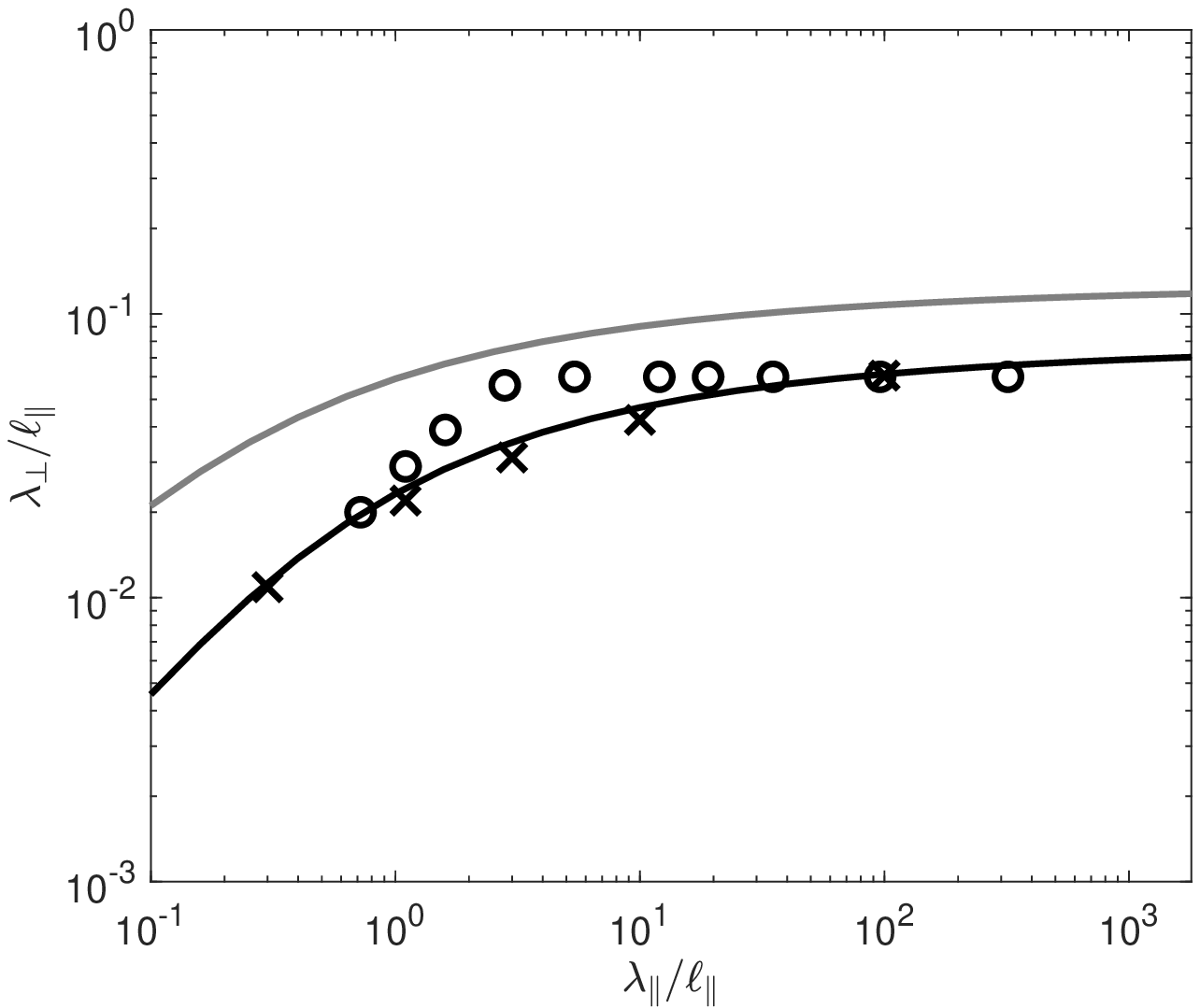}
\end{minipage}
\hfill
\begin{minipage}[b]{0.48\textwidth}
\includegraphics[width=\textwidth]{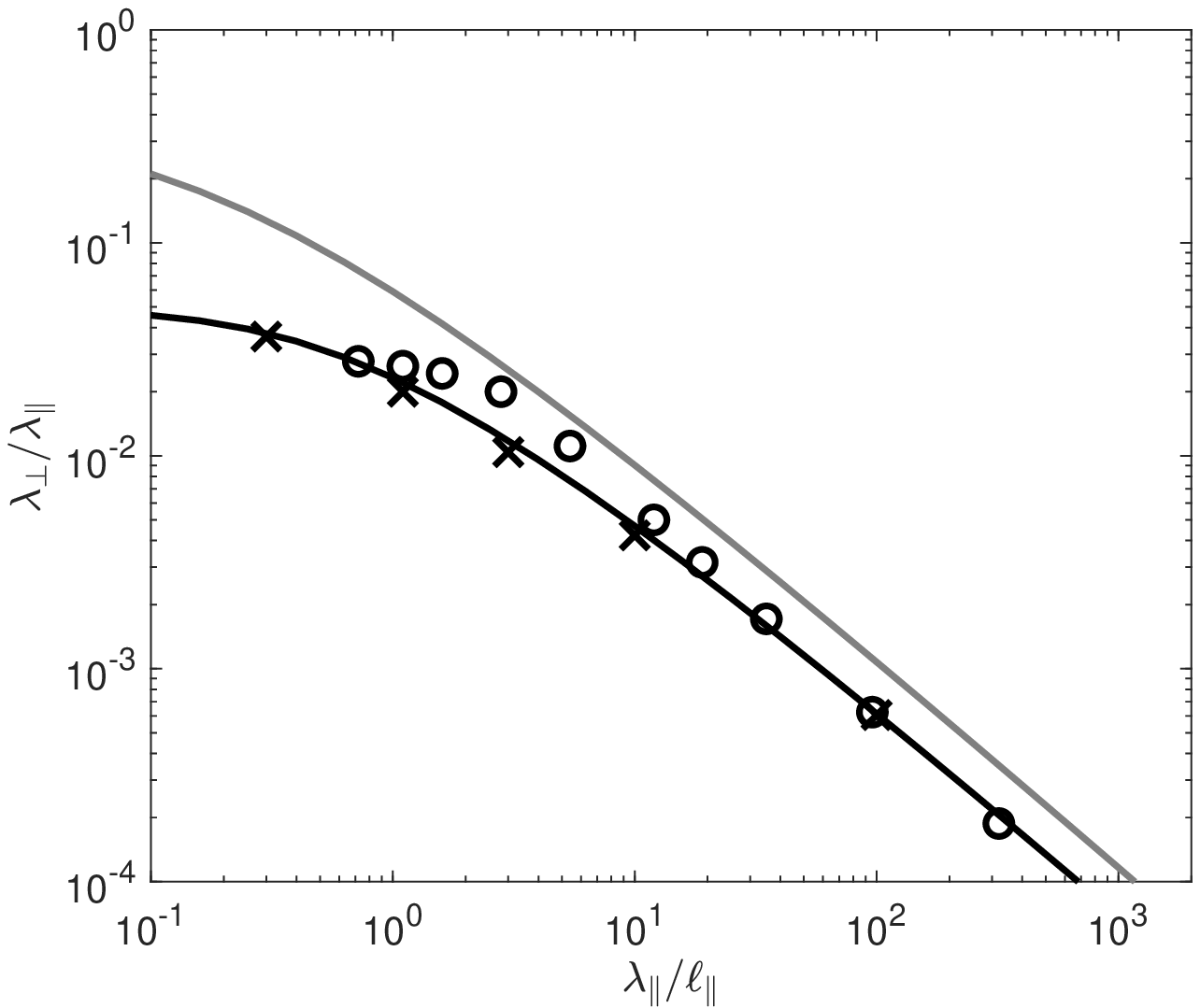}
\end{minipage}
\caption{The left panel shows the perpendicular mean free path versus the parallel mean free path and the right panel shows the ratio
$\lambda_{\perp} / \lambda_{\parallel} \equiv \kappa_{\perp} / \kappa_{\parallel}$ versus the parallel mean free path. The black solid
lines correspond to the diffusive FLPD results as given by Eq. (\ref{dimlessEUNLT}). The circles represent the test-particle simulations
published in Qin \& Shalchi (2012). The crosses represent time-dependent FLPD theory as given by Eq. (\ref{holyequation}) or, in code units,
by Eq. (\ref{slab2Dfornumerics}). For comparison we have also shown the NLGC/UNLT results without correction factor, meaning that we have
set $a^2 = 1$ (gray solid line). To obtain the visualized results the following parameter values have been used: $\delta B^2 / B_0^2 = 1$,
$\ell_{\perp} = 0.1 \ell_{\parallel}$, $q=1.5$, and a slab fraction of $20 \%$.}
\label{SimComp6}
\end{figure*}

In Fig. \ref{SimComp1}-\ref{SimComp6} we compare the FLPD theory developed in the current paper with previously performed test-particle simulations.
Those simulations were performed for a two-component turbulence model and were taken from Hussein et al. (2015), Qin \& Shalchi (2012), as well as
Arendt \& Shalchi (2020). The simulations are compared with the diffusive version of FLPD theory, as given by Eq. (\ref{diffEUNLT}), as well as the
time-dependent version given by Eq. (\ref{holyequation}). For the latter case the reader can find the corresponding mathematical details in
the Appendix B of the current paper (see Eq. (\ref{slab2Dfornumerics})). The used parameter values can be found in the caption of the corresponding
figure.

We have covered a large parameter space meaning that we have changed turbulence strength $\delta B / B_0$, the scale ratio $\ell_{\perp}/\ell_{\parallel}$,
the slab fraction, and even the energy range spectral index $q$. One can easily see that the agreement between simulations and FLPD theory is remarkable
in all cases. A correction parameter $a^2$ is no longer needed in order to achieve this agreement nor does FLPD theory contain any free parameter (see
Table \ref{Agreements} for a comparison of the different theories). Of course FLPD theory without diffusion approximation is more accurate but even if
a diffusion approximation is used, the agreement is more than satisfactory. This is important because using the diffusion approximation leads to a strong
mathematical simplification of our new theory.

\begin{table*}
\caption{The different theories and their agreement with heuristic arguments and test-particle simulations. Compared are the Non-Linear Guiding
Center (NLGC) theory of Matthaeus et al. (2003), the Unified Non-Linear Transport (UNLT) theory, as well as the Field Line - Particle Decorrelation (FLPD)
theory developed in the current article. Simple concepts such as quasi-linear theory or hard sphere scattering disagree with simulations
and heuristic arguments.}
\begin{center}
\begin{tabular}{|l|l|l|l|}
\hline
Theory				& Agreement with  				& Agreement with Simulations 		& Agreement with Simulations 			\\
					& Heuristic Arguments			& for Small Kubo Number Turbulence	& for Large Kubo Number Turbulence		\\[0.1cm]
\hline
NLGC Theory			& NO							& NO								& Only with Correction Factor $a^2$		\\[0.1cm]
UNLT Theory			& Only for small Kubo numbers	& YES								& Only with Correction Factor $a^2$		\\[0.1cm]
FLPD Theory			& YES							& YES								& YES									\\[0.1cm]
\hline
\end{tabular}
\end{center}
\label{Agreements}
\end{table*}

\begin{figure*}[!tbp]
\centering
\begin{minipage}[b]{0.48\textwidth}
\includegraphics[width=\textwidth]{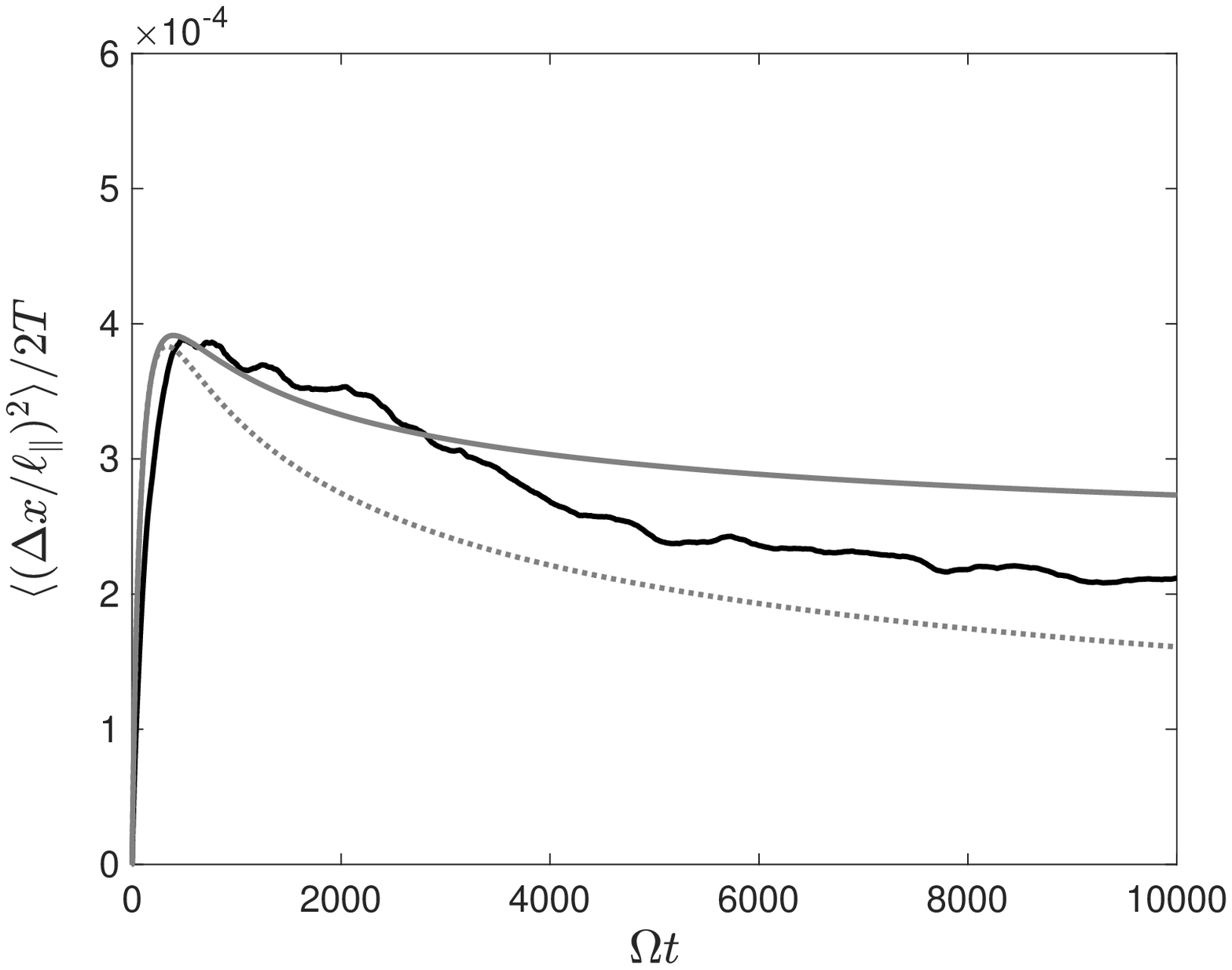}
\end{minipage}
\hfill
\begin{minipage}[b]{0.48\textwidth}
\includegraphics[width=\textwidth]{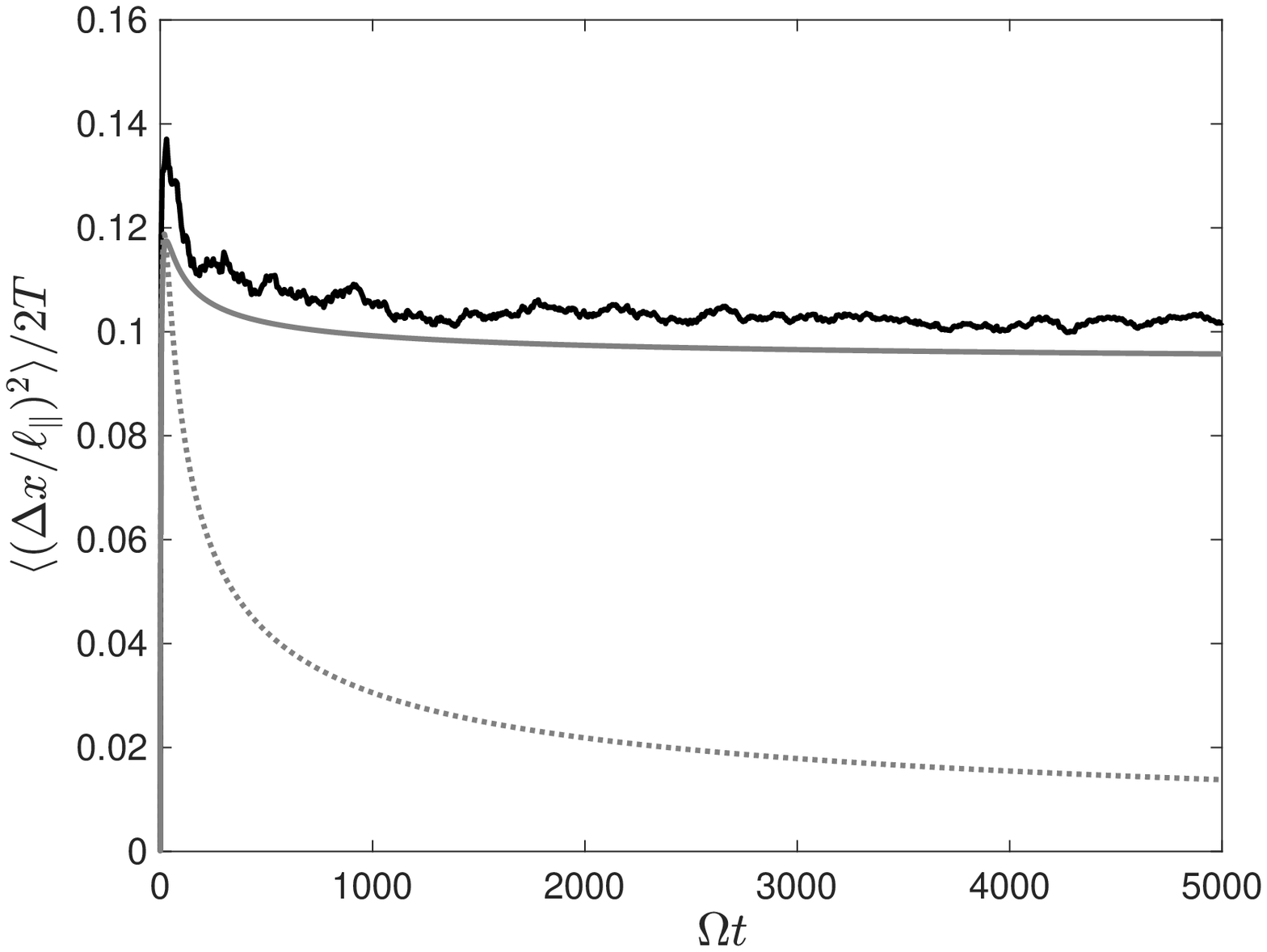}
\end{minipage}
\caption{Shown are time-dependent perpendicular diffusion coefficients. Here we have used the dimensionless time $T = \Omega t$ and the
gyro-frequency $\Omega$. Compared are the test-particle simulations published in Arendt \& Shalchi (2020) represented by the black solid lines and the
time-dependent FLPD theory represented by the gray solid lines. The latter results were obtained by solving Eq. (\ref{slab2Dfornumerics}) numerically.
The used parameter values are $q=3$, $\ell_{\parallel} = \ell_{\perp}$, $\delta B_{slab}^2 / B_0^2 = 0.20$, and $\delta B_{2D}^2 / B_0^2 = 0.80$.
For comparison we have also shown time-dependent FLPD theory without transverse complexity factor corresponding to compound sub-diffusion (gray dotted lines).
The left panel shows the perpendicular diffusion coefficient for lower particle energies ($R=v/(\Omega \ell_{\parallel})=0.01$) corresponding to a parallel
mean free path of $\lambda_{\parallel} / \ell_{\parallel} = 0.954$. The right panel shows results for intermediate particle energies ($R=v/(\Omega \ell_{\parallel})=1$)
corresponding to a parallel mean free path of $\lambda_{\parallel} / \ell_{\parallel} = 17.1$.}
\label{TimePlot}
\end{figure*}

Fig. \ref{TimePlot} shows as an example the time-dependent perpendicular diffusion coefficient for two different sets of particle and turbulence
parameters. Very nicely we can see that the time-dependent version of FLPD theory provides the initial ballistic regime and thereafter the
transport becomes weakly sub-diffusive. For later times diffusion is restored because the transverse complexity of the turbulence influences the
particle motion. The sequence \textit{ballistic transport $\rightarrow$ sub-diffusion $\rightarrow$ normal diffusion} is more pronounced for shorter
parallel mean free paths corresponding to lower particle rigidities.

\section{Summary and Conclusion}

Particle transport and perpendicular diffusion in particular are fundamental problems in modern physics. To explore how particles move across
a mean magnetic field is explored since more than half a century and started with the work of Jokipii (1966). For decades it was
unclear how perpendicular transport in space and astrophysical plasmas works. Significant progress has been made by the famous paper of
Rochester \& Rosenbluth (1978) but their work focused on laboratory plasmas where collisions play a major role. Over years several attempts
have been made to solve this problem in space and astrophysics often based on simple concepts such as quasi-linear theory or hard sphere
scattering. However, all those approaches fail in almost all cases. A major breakthrough has been made by Matthaeus et al. (2003) where
a non-linear integral equation has been derived based on a set of approximations. The latter theory, commonly called the \textit{Non-Linear
Guiding Center (NLGC) theory}, was the first approach in history which showed agreement with simulations and observations. However, this
approach was not the final solution to the problem, mainly because of two reasons. First, NLGC theory provides a non-vanishing diffusion
coefficient for slab turbulence but it is known for sure that perpendicular transport in turbulence without transverse structure has to be
sub-diffusive since particles are tied to magnetic field lines forever. Second, for other turbulence configurations, the theory provides a
result which is typically a factor three too large. Therefore, Matthaeus et al. (2003) incorporated a correction factor $a^2$ to achieve
agreement with test-particle simulations.

In Shalchi (2010) the so-called \textit{diffusive unified non-linear transport (UNLT) theory} was derived systematically (see Eq. (\ref{UNLT})
of the current paper). Diffusive UNLT theory provides $\kappa_{\perp} = 0$ for slab turbulence resolving the first of the two problems listed above.
Furthermore, the theory contains the field line random walk theory developed by Matthaeus et al. (1995) as special limit and can, therefore,
be seen as a unified theory for particles and magnetic field lines. A time-dependent version of UNLT theory (see Eq. (\ref{Generalsigmap})
of the current paper) was developed by Shalchi (2017) and Lasuik \& Shalchi (2017). The latter theory allows for a description of perpendicular
transport in all regimes including the initial ballistic regime, the sub-diffusive regime, and the normal diffusive regime. The latter case is
obtained due to the impact the transverse complexity of the turbulence has. However, diffusive as well as time-dependent UNLT theories
did not solve the second problem listed above. Both theories require to use the correction parameter $a^2$ in the case of large Kubo number
turbulence including quasi-2D turbulence and two-component turbulence.

Independent of the development of analytical theories and numerical tools, there was always a desire to understand perpendicular
transport in space plasmas in a more heuristic way. By following the spirit of Rochester \& Rosenbluth (1978), a heuristic approach
was finally developed by Shalchi (2019a) and Shalchi (2020b). This heuristic description provides simple analytical forms of the
perpendicular diffusion coefficient in certain limits. Those forms include the traditional field line random walk limit (see
Eq. (\ref{FLRWlimit})), the so-called fluid limit (see Eq. (\ref{Fluid})), as well as a collisionless version of Rochester \& Rosenbluth (1978),
short \textit{CLRR limit} (see Eq. (\ref{CLRR}) of the current paper). Although it was not clear from this heuristic description whether
one finds the fluid or the CLRR limit for short parallel mean free paths and large Kubo number turbulence, the heuristic approach provided
a possible explanation why $a^2$ is needed and why one needs to set $a^2 = 1/3$.

Heuristic arguments are important in order to understand the underlying physics of particle transport. However, one still needs an
analytical theory in order to describe perpendicular transport systematically. Such a systematic theory was finally developed in the
current paper. The new theory can be understood as a generalized compound sub-diffusion approach where a transverse complexity factor
is included allowing the particles to decorrelate from the original magnetic field line they were tied to. Thus, we call the new
approach \textit{Field Line - Particle Decorrelation (FLPD) theory}. We derived a time-dependent version (see Eq. (\ref{holyequation}))
as well as a diffusive version (see Eq. (\ref{diffEUNLT})) of this theory. FLPD theory has several important features. For small Kubo
number turbulence the new theory becomes identical compared to UNLT theory (see Fig. \ref{TheDiagram} for a detailed explanation of 
the relations between all those different approaches). For slab turbulence we find compound sub-diffusion as the final state of the transport.
For non-slab turbulence we find compound sub-diffusion after the initial ballistic regime as long as transverse complexity in not
yet relevant. Thereafter diffusion is restored and this happens as soon as transverse complexity becomes relevant (see, e.g., Fig. \ref{TimePlot}
of the current paper). Both, the fluid limit as well as the CLRR limit are contained in the new theory. For large Kubo number and quasi-2D
turbulence, we find in the diffusion limit Eq. (\ref{kperpg2D2}). Compared to the theories of Matthaeus et al. (2003) and Shalchi (2010),
we find an extra term in the denominator of the fraction in the wave number integral. This extra term corresponds to a CLRR diffusion
term and provides the final piece of the puzzle which was missing for almost two decades. Fig. \ref{EUNLTRegimes} explains our current
understanding of perpendicular transport in three-dimensional turbulence. Fig. \ref{TheDiagram} compares different analytical theories
developed previously and Table \ref{Agreements} explains when they agree with heuristic arguments and simulations.

Time-dependent and diffusive versions of the new theory can be combined with a two-component turbulence model (see, e.g., Eq. (\ref{slab2Dfornumerics})
of the current paper). This is an important example because solar wind turbulence can be well approximated by this type
of model (see, e.g., Matthaeus et al. (1990)). For this specific turbulence approximation we compared the new theory with previously
performed test-particle simulations published in Hussein et al. (2015), Qin \& Shalchi (2012), and Arendt \& Shalchi (2020).
These sets of simulations were performed over the past few years by using four different test-particle codes. The comparison is
visualized in Figs. \ref{SimComp1}-\ref{SimComp6} as well as Fig. \ref{TimePlot}. Very clearly one finds a remarkable agreement
between the new theory and the simulations regardless whether the diffusion approximation is used or not. It needs to be emphasized
that the new theory does not contain any correction parameter $a^2$ or any other type of free parameter.

\begin{acknowledgements}
\textit{Support by the Natural Sciences and Engineering Research Council (NSERC) of Canada is acknowledged. Furthermore, I would like to thank
G. P. Zank for some very useful comments concerning magnetic turbulence in the solar wind and the very local interstellar medium.}
\end{acknowledgements}

\onecolumngrid
\appendix
\section{The Particle Velocity Correlation Function}

Our aim is to develop a systematic theory for the perpendicular particle diffusion coefficient and, more general, mean square displacements
of particle orbits. In the following we focus on the perpendicular velocity correlation function $\langle V_x (t) V_x (0) \rangle$. Instead
of using Eq. (\ref{chapman}), we start our derivation with
\be
\Big< V_x (t) V_x (0) \Big> = \frac{v^2}{3 B_0^2} \int_{-\infty}^{+\infty} d z \;
\Big< \delta B_x \left[ \vec{x} (z) \right] \delta B_x \left[ \vec{x} (0) \right] \Big> f_{\parallel} \big( z; t \big).
\label{chapmancorr}
\ee
The factor $v^2 / 3$ therein is needed to satisfy the assumption of isotropic initial conditions. Note, the magnetic field correlations have to be
computed along the magnetic field line $\vec{x} (z)$. It follows from the TGK formula (\ref{tgk}) that velocity correlations and means square displacement
are related via Eq. (\ref{relateMSDandVCF}). The same can be assumed for magnetic field lines where we have
\be
\frac{d^2}{d z^2} \Big< (\Delta x)^2 \Big>_{FL} = \frac{2}{B_0^2} \Big< \delta B_x \left[ \vec{x} (z) \right] \delta B_x \left[ \vec{x} (0) \right] \Big>.
\ee
Therefore, Eq. (\ref{chapmancorr}) can be written as
\be
\frac{d^2}{d t^2} \Big< (\Delta x)^2 \Big>_P = \int_{-\infty}^{+\infty} d z \; \left[ \frac{d^2}{d z^2} \Big< (\Delta x)^2 \Big>_{FL} \right]
f_{\parallel} \big( z; t \big).
\label{MSDequation}
\ee
Shalchi \& Kourakis (2007b) derived Eq. (\ref{Kourakis2007}) for the field line mean square displacement. Using this in Eq. (\ref{MSDequation}) yields
\be
\frac{d^2}{d t^2} \Big< (\Delta x)^2 \Big>_P = \frac{2}{B_0^2} \int d^3 k \; P_{xx} \big( \vec{k} \big)
\int_{-\infty}^{+\infty} d z \; \cos \big( k_{\parallel} z \big)
e^{- \frac{1}{2} \langle \left( \Delta x \right)^2 \rangle_{FL} k_{\perp}^2} f_{\parallel} \big( z; t \big).
\ee
For the magnetic field lines we now employ a diffusion approximation of the form $\langle \left( \Delta x \right)^2 \rangle_{FL} = 2 \kappafl \left| z \right|$
and we find
\be
\frac{d^2}{d t^2} \Big< (\Delta x)^2 \Big>_P = \frac{2}{B_0^2} \int d^3 k \; P_{xx} \big( \vec{k} \big)
\int_{-\infty}^{+\infty} d z \; f_{\parallel} \big( z; t \big) \cos \big( k_{\parallel} z \big)
e^{- \kappafl k_{\perp}^2 \left| z \right|}.
\label{Bstep4}
\ee
The next step is to replace the parallel distribution function of the particle $f_{\parallel} \left( z; t \right)$. First we use the Fourier
representation given by Eq. (\ref{fparallelFourier}). Using this in Eq. (\ref{Bstep4}) yields
\be
\frac{d^2}{d t^2} \Big< (\Delta x)^2 \Big>_P = \frac{2 v^2}{3 B_0^2} \int_{-\infty}^{+\infty} d k_{\parallel}^{\prime} \; F \big( k_{\parallel}^{\prime}, t \big)
\int d^3 k \; P_{xx} \big( \vec{k} \big) \int_{-\infty}^{+\infty} d z \; \cos \big( k_{\parallel} z \big)
e^{- \kappafl k_{\perp}^2 \left| z \right| + i k_{\parallel}^{\prime} z}.
\label{Bstep5}
\ee
The $z$-integral therein can be solved via Eq. (\ref{beforeresonance}). Furthermore, we use the resonance function defined via Eq. (\ref{resonancefunction}). Therewith we can write Eq. (\ref{Bstep5}) as
\be
\frac{d^2}{d t^2} \Big< (\Delta x)^2 \Big>_P = \frac{2 v^2}{3 B_0^2} \int_{-\infty}^{+\infty} d k_{\parallel}^{\prime} \; F \left( k_{\parallel}^{\prime}, t \right)
\int d^3 k \; P_{xx} \big( \vec{k} \big) \left[ R_{+} \big( \vec{k}, k_{\parallel}^{\prime} \big) + R_{-} \big( \vec{k}, k_{\parallel}^{\prime} \big) \right].
\label{Bstep6}
\ee
For the parallel distribution function in Fourier space we can employ Eq. (\ref{formforW}) where the two parameter $\omega_+$ and $\omega_-$ are given by
Eq. (\ref{omegapm}). Comparing this with Eq. (\ref{formforWoldapp}) allows us to write this as
\be
F \left( k_{\parallel}, t \right) = \frac{1}{2 \pi} \frac{3}{v^2} \xi \left( k_{\parallel}, t \right).
\ee
Using this in Eq. (\ref{Bstep6}) yields
\be
\frac{d^2}{d t^2} \Big< (\Delta x)^2 \Big>_P = \frac{1}{\pi B_0^2} \int_{-\infty}^{+\infty} d k_{\parallel}^{\prime} \; \xi \left( k_{\parallel}^{\prime}, t \right)
\int d^3 k \; P_{xx} \big( \vec{k} \big) \left[ R_{+} \big( \vec{k}, k_{\parallel}^{\prime} \big) + R_{-} \big( \vec{k}, k_{\parallel}^{\prime} \big) \right]
\label{altderiv}
\ee
in perfect agreement with Eq. (\ref{anotherstep6}).

\section{Time-Dependent Results for Two-Component Turbulence}

In order to achieve an analytical description of perpendicular transport with high accuracy we need to employ the time-dependent version of FLPD theory
meaning that we do not use a diffusion approximation. The corresponding equation is given by (see Eq. (\ref{holyequation}) of the main part of this
paper)
\be
\frac{d^2}{d t^2} \Big< ( \Delta x )^2 \Big>
= \frac{1}{\pi B_0^2} \int_{-\infty}^{+\infty} d k_{\parallel}^{\prime} \; \xi \big( k_{\parallel}^{\prime}, t \big)
\int d^3 k \; P_{xx} \big( \vec{k} \big) \left[ R_{+} \big( \vec{k}, k_{\parallel}^{\prime} \big) + R_{-} \big( \vec{k}, k_{\parallel}^{\prime} \big) \right]
e^{- \frac{1}{2} \langle ( \Delta x )^2 \rangle k_{\perp}^2}.
\label{copyholy}
\ee
In the following we combine Eq. (\ref{copyholy}) with the two-component turbulence model. Therefore, there will be two contributions
to the spectral tensor and Eq. (\ref{copyholy}) will consist of two contributions as well. As pointed out in the main part of this paper,
the slab contribution provided by the new theory is in perfect coincidence with the time-dependent UNLT result. For the two-dimensional
contribution, on the other hand, we can employ Eqs. (\ref{timedependentfor2D}) and (\ref{2Dresonancefunction}). Therefore, Eq. (\ref{copyholy})
becomes for two-component turbulence
\bdm
\frac{d^2}{d t^2} \Big< ( \Delta x )^2 \Big>
& = & \frac{8 \pi}{B_0^2} \int_{0}^{\infty} d k_{\parallel} \; g^{slab} \bigl( k_{\parallel} \bigr) \xi \left( k_{\parallel}, t \right) \nonumber\\
& + & \frac{4}{B_0^2} \int_{0}^{\infty} d k_{\parallel} \; \xi \left( k_{\parallel}, t \right) \int_{0}^{\infty} d k_{\perp} \; g^{2D} (k_{\perp}) \frac{\kappafl k_{\perp}^{2}}{k_{\parallel}^2 + \left( \kappafl k_{\perp}^{2} \right)^2} e^{- \frac{1}{2} \langle ( \Delta x )^2 \rangle k_{\perp}^2}.
\label{diffeqforslab2D}
\edm
Therein we need to replace three functions, namely the slab spectrum $g^{slab} \big( k_{\parallel} \big)$, the 2D spectrum $g^{2D} (k_{\perp})$,
as well as the parallel correlation function $\xi \left( k_{\parallel}, t \right)$. For the slab spectrum we employ Eq. (\ref{SlabSpectrum}) and
the 2D spectrum and the parallel correlation function are given by Eqs. (\ref{2dspec}) and (\ref{formforWoldapp}), respectively. Therewith our
differential equation (\ref{diffeqforslab2D}) becomes
\bdm
\frac{d^2}{d t^2} \Big< ( \Delta x )^2 \Big>
& = & 4 C(s) \frac{v^2}{3} \frac{\delta B_{slab}^2}{B_0^2} \ell_\parallel \int_{0}^{\infty} d k_{\parallel} \;
\frac{1}{\big[ 1 + \left( k_\parallel \ell_\parallel \right)^2 \big]^{s/2}}
\frac{1}{\omega_+ - \omega_-} \big[ \omega_+ e^{\omega_+ t} - \omega_- e^{\omega_- t} \big] \nonumber\\
& + & \frac{8 D(s, q)}{\pi} \frac{v^2}{3} \frac{\delta B_{2D}^2}{B_0^2} \ell_{\perp} \int_{0}^{\infty} d k_{\parallel} \; \frac{1}{\omega_+ - \omega_-}
\big[ \omega_+ e^{\omega_+ t} - \omega_- e^{\omega_- t} \big] \nonumber\\
& \times & \int_{0}^{\infty} d k_{\perp} \; \frac{(k_{\perp} \ell_{\perp})^{q}}{\big[ 1 + \big( k_{\perp} \ell_{\perp} \big)^2 \big]^{(s+q)/2}}
\frac{\kappafl k_{\perp}^{2}}{k_{\parallel}^2 + \left( \kappafl k_{\perp}^{2} \right)^2} e^{- \frac{1}{2} \langle ( \Delta x )^2 \rangle k_{\perp}^2}.
\edm
In the latter equation we have used the parameters $\omega_+$ and $\omega_-$ given by Eq. (\ref{omegapm}). In the slab contribution we employ the
integral transformation $x = k_{\parallel} \ell_{\parallel}$ and in the 2D contribution the transformation $y = k_{\perp} \ell_{\perp}$ to write
\bdm
\frac{d^2}{d t^2} \Big< ( \Delta x )^2 \Big>
& = & 4 C(s) \frac{v^2}{3} \frac{\delta B_{slab}^2}{B_0^2} \int_{0}^{\infty} d x \;
\frac{1}{\big[ 1 + x^2 \big]^{s/2}}
\frac{1}{\omega_+ - \omega_-} \big[ \omega_+ e^{\omega_+ t} - \omega_- e^{\omega_- t} \big] \nonumber\\
& + & \frac{8 D(s, q)}{\pi} \frac{v^2}{3} \frac{\delta B_{2D}^2}{B_0^2} \ell_{\perp} \int_{0}^{\infty} d k_{\parallel} \; \frac{1}{\omega_+ - \omega_-}
\left[ \omega_+ e^{\omega_+ t} - \omega_- e^{\omega_- t} \right] \nonumber\\
& \times & \int_{0}^{\infty} d y \; \frac{y^q}{\left[ 1 + y^2 \right]^{(s+q)/2}}
\frac{\kappafl/\ell_{\perp} y^{2}}{(\ell_{\perp} k_{\parallel})^2 + \left( \kappafl/\ell_{\perp} y^2 \right)^2}
e^{- \langle ( \Delta x )^2 \rangle y^2 / (2 \ell_{\perp}^2)}.
\edm
To continue we use the dimensionless time and dimensionless mean square displacement $\tau := v t / \ell_{\parallel}$ and
$\sigma := \langle ( \Delta x )^2 \rangle / \ell_{\perp}^2$, respectively. Therewith we can write
\bdm
\frac{d^2 \sigma}{d \tau^2}
& = & \frac{4}{3} C(s) \frac{\ell_{\parallel}^2}{\ell_{\perp}^2} \frac{\delta B_{slab}^2}{B_0^2} \int_{0}^{\infty} d x \;
\frac{1}{\left[ 1 + x^2 \right]^{s/2}} S \big( \omega_+, \omega_-, t \big) \nonumber\\
& + & \frac{8}{3 \pi} D(s, q) \frac{\ell_{\parallel}^2}{\ell_{\perp}^2} \frac{\delta B_{2D}^2}{B_0^2} \ell_{\perp}
\int_{0}^{\infty} d k_{\parallel} \; S \big( \omega_+, \omega_-, t \big) \nonumber\\
& \times & \int_{0}^{\infty} d y \; \frac{y^q}{\left[ 1 + y^2 \right]^{(s+q)/2}}
\frac{\kappafl/\ell_{\perp} y^{2}}{(\ell_{\perp} k_{\parallel})^2 + \left( \kappafl/\ell_{\perp} y^2 \right)^2} e^{- \frac{1}{2} \sigma y^2}
\label{stillkparainit}
\edm
where we have used
\be
S \big( \omega_+, \omega_-, x, t \big) = \frac{1}{\omega_+ - \omega_-} \left[ \omega_+ e^{\omega_+ t} - \omega_- e^{\omega_- t} \right].
\ee
The two parameters $\omega_+$ and $\omega_-$ can be written as
\be
\omega_{\pm} = \frac{v}{\ell_{\parallel}} \left[ - \frac{\ell_{\parallel}}{2 \lambda_{\parallel}} \pm \sqrt{\left( \frac{\ell_{\parallel}}{2 \lambda_{\parallel}} \right)^2 - \frac{1}{3} x^2} \right]
\ee
so that
\be
\omega_{\pm} t = \left[ - \frac{\ell_{\parallel}}{2 \lambda_{\parallel}} \pm \sqrt{\left( \frac{\ell_{\parallel}}{2 \lambda_{\parallel}} \right)^2 - \frac{1}{3} x^2} \right] \tau.
\ee
Using the integral transformation $x = k_{\parallel} \ell_{\parallel}$ in the remaining $k_{\parallel}$-integral allows us to write
Eq. (\ref{stillkparainit}) as
\bdm
\frac{d^2 \sigma}{d \tau^2}
& = & \frac{4}{3} C(s) \frac{\ell_{\parallel}^2}{\ell_{\perp}^2} \frac{\delta B_{slab}^2}{B_0^2} \int_{0}^{\infty} d x \;
\frac{1}{\left[ 1 + x^2 \right]^{s/2}} S \big( \omega_+, \omega_-, t \big) \nonumber\\
& + & \frac{8}{3 \pi} D(s, q) \frac{\ell_{\parallel}}{\ell_{\perp}} \frac{\delta B_{2D}^2}{B_0^2}
\int_{0}^{\infty} d x \; S \big( \omega_+, \omega_-, t \big) \nonumber\\
& \times & \int_{0}^{\infty} d y \; \frac{y^q}{\left[ 1 + y^2 \right]^{(s+q)/2}}
\frac{\kappafl/\ell_{\perp} y^{2}}{\ell_{\perp}^2 x^2 / \ell_{\parallel}^2 + \left( \kappafl/\ell_{\perp} y^2 \right)^2} e^{- \frac{1}{2} \sigma y^2}.
\label{slab2Dfornumerics}
\edm
The latter formula is a second-order differential equation for the function $\sigma$. Since all quantities therein are dimensionless, this differential
equation can be solved numerically. This was done in the current paper to produce the crosses in Fig. \ref{SimComp1}-\ref{SimComp6} as well as the
time-dependent plots in Fig. \ref{TimePlot}. It has to be noted that the first term in Eq. (\ref{slab2Dfornumerics}), which is coming from the slab modes,
is sub-diffusive. We still kept it to obtain a more accurate result especially for earlier times.

{}

\end{document}